\pdfoutput=1

\documentclass[preprint,authoryear,5p]{elsarticle}

\usepackage{paralist}
\usepackage{psfrag,color}
\usepackage{graphicx,amssymb}
\usepackage{epsf,psfig}
\usepackage[latin1]{inputenc}
\usepackage{natbib}
\usepackage[unicode]{hyperref}

\journal{New Astronomy}

\begin{document}

\begin{frontmatter}
\title{Escape dynamics in a binary system of interacting galaxies}

\author[]{Euaggelos E. Zotos\corref{cor}}
\ead{evzotos@physics.auth.gr}

\cortext[cor]{Corresponding author}

\address[]{Department of Physics, School of Science, \\
Aristotle University of Thessaloniki, \\
GR-541 24, Thessaloniki, Greece}

\begin{abstract}
The escape dynamics in an analytical gravitational model which describes the motion of stars in a binary system of interacting dwarf spheroidal galaxies is investigated in detail. We conduct a numerical analysis distinguishing between regular and chaotic orbits as well as between trapped and escaping orbits, considering only unbounded motion for several energy levels. In order to distinguish safely and with certainty between ordered and chaotic motion, we apply the Smaller ALingment Index (SALI) method. It is of particular interest to locate the escape basins through the openings around the collinear Lagrangian points $L_1$ and $L_2$ and relate them with the corresponding spatial distribution of the escape times of the orbits. Our exploration takes place both in the configuration $(x,y)$ and in the phase $(x,\dot{x})$ space in order to elucidate the escape process as well as the overall orbital properties of the galactic system. Our numerical analysis reveals the strong dependence of the properties of the considered escape basins with the total orbital energy, with a remarkable presence of fractal basin boundaries along all the escape regimes. It was also observed, that for energy levels close to the critical escape energy the escape rates of orbits are large, while for much higher values of energy most of the orbits have low escape periods or they escape immediately to infinity. We hope our outcomes to be useful for a further understanding of the escape mechanism in binary galaxy models.
\end{abstract}


\end{frontmatter}

\section{Introduction}
\label{intro}

Over the years many studies have been devoted on the issue of escaping particles from dynamical systems. Especially the issue of escapes in Hamiltonian systems is directly related to the problem of chaotic scattering which has been an active field of research over the last decades and it still remains open \citep[e.g.,][]{BGOB88,JS88,CK92,BST98,ML02,SASL06,SSL07,SS08,SHSL09,SS10}. The problem of escape is a classical problem in simple Hamiltonian nonlinear systems \citep[e.g.,][]{AVS01,AS03,AVS09,BBS08,BSBS12,Z14a} as well as in dynamical astronomy \citep[e.g.,][]{HB83,BTS96,BST98,dML00,Z12a}. Escaping orbits in the classical Restricted Three-Body Problem (RTBP) is another typical example \citep[e.g.,][]{N04,N05,dAT14}.

Nevertheless, the issue of escaping orbits in Hamiltonian systems is by far less explored than the closely related problem of chaotic scattering. In this situation, a test particle coming from infinity approaches and then scatters off a complex potential. This phenomenon is well investigated as well interpreted from the viewpoint of chaos theory \citep[e.g.,][]{BGOB88,BGO90,BOG89,J87,JLS99,JMS95,JP89,JR90,JS87,JT91,LMG00,LGB93,LFO91,LJ99}.

In open Hamiltonian systems an issue of great importance is the determination of the basins of escape, similar to basins of attraction in dissipative systems or even the Newton-Raphson fractal structures. An escape basin is defined as a local set of initial conditions of orbits for which the test particles escape through a certain exit in the equipotential surface for energies above the escape value. Basins of escape have been studied in many earlier papers (e.g., \citep{BGOB88,C02,KY91,PCOG96}). The reader can find more details regarding basins of escape in \citep{C02}, while the review \citep{Z14b} provides information about the escape properties of orbits in a multi-channel dynamical system composed of a two-dimensional perturbed harmonic oscillator. The boundaries of an escape basins may be fractal (e.g., \citep{AVS09,BGOB88}) or even respect the more restrictive Wada property (e.g., \citep{AVS01}), in the case where three or more escape channels coexist in the equipotential surface.

Escaping and trapped motion of stars in stellar systems is an another issue of great importance. In a previous article \citep{Z12a}, we explored the nature of orbits of stars in a galactic-type potential, which can be considered to describe local motion in the meridional $(R,z)$ plane near the central parts of an axially symmetric galaxy. It was observed, that apart from the trapped orbits there are two types of escaping orbits, those which escape fast and those which need to spend vast time intervals inside the equipotential surface before they find the exit and eventually escape. The escape dynamics and the dissolution process of a star cluster embedded in the tidal field of a parent galaxy was investigated in \citet{EJSP08}. Conducting a scanning of the available phase space the authors managed to obtain the basins of escape and the respective escape rates of the orbits, revealing that the higher escape times correspond to initial conditions of orbits near the fractal basin boundaries. The investigation was expanded into three dimensions in \citet{Z15} where we revealed the escape mechanism of three-dimensional orbits in a tidally limited star cluster. Furthermore, \citet{EP14} explored the escape dynamics in the close vicinity of and within the critical area in a two-dimensional barred galaxy potential, identifying the escape basins both in the phase and the configuration space.

The numerical approach of the above-mentioned papers serves as the basis of this work. The main objective of our numerical exploration is to determine which orbits escape and which remain trapped, distinguishing simultaneously between regular and chaotic trapped motion. Furthermore, we shall try to locate the escape basins which reflect the orbital structure of the system and they also determine through which channel the orbit escape to infinity. To our knowledge, this is the first that the escape dynamics of a binary system of interacting galaxies is numerically investigated. Our work is quite similar to \citet{Z15} where we studied the escape process in a star cluster rotating around its parent galaxy. In \citet{Z15} however, the dynamical system had three degrees of freedom (3D), while the present one is only two-dimensional (2D).

The article is organized as follows: In Section \ref{galmod} we present in detail the structure and the properties of our binary galaxy model. All the computational methods we used in order to determine the nature of orbits are described in Section \ref{cometh}. In the following Section, we conduct a systematic numerical investigation revealing the overall orbital structure (bounded regions and basins of escape) of the binary galaxy and showing how it is affected by the total orbital energy. Our paper ends with Section \ref{disc}, where the discussion and the main conclusions are given.

\section{The binary galaxy model}
\label{galmod}

The aim of this research is to explore the properties of motion in the planar softened circular restricted three-body problem. Our analytic gravitational model consists of a pair of dwarf spheroidal galaxies. The two spheroidal galaxies move in circular orbits around their common center of gravity, which is assumed to be fixed at the origin $(0,0)$ of the coordinates. The third body (a star test particle) moves in the same plane under the gravitational field of the two galaxies. As a first step we shall consider the case where the two galaxies are identical (same mass, same structure) similarly to the Copenhagen case of the classical RTBP.

To model the dynamical properties of the spheroidal galaxies we use the well known spherically symmetric Plummer potential \citep{P11}. Therefore, the potential which describes the motion around the first galaxy (hereafter galaxy $\rm G_1$) is given by the equation
\begin{equation}
\Phi_1(x,y) = \frac{- \ G M_1}{\sqrt{R^2 + c_1^2}},
\label{pot1}
\end{equation}
where $R^2 = x^2 + y^2$, while $M_1$ is the mass and $c_1$ the core radius of galaxy $\rm G_1$. Similarly, galaxy $\rm G_2$ is described by the potential
\begin{equation}
\Phi_2(x,y) = \frac{- \ G M_2}{\sqrt{R^2 + c_2^2}},
\label{pot2}
\end{equation}
where $M_2$ is the mass and $c_2$ the core radius of galaxy $\rm G_2$. The core radius $c_i$ acts as a softening parameter which eliminates the problem of critical collision orbits which is present in the classical RTBP.

We shall apply the theory of the softened circular restricted three-body problem. The two galaxies move in circular orbits in an inertial frame OXYZ with the origin at the center of mass of the system with a constant angular velocity $\Omega_{\rm p} > 0$, given by Kepler's third law
\begin{equation}
\Omega_{\rm p}= \sqrt{\frac{G M_t}{d^3}},
\label{omega}
\end{equation}
where $M_{\rm t} = M_1 + M_2$ is the total mass of the system, while $d$ is the distance between the centers of the two bodies. A clockwise, rotating frame Oxyz, is used with the axis Oz coinciding with the axis OZ and the axis Ox coinciding with the straight line joining the centers of two galaxies. In this frame, which rotates with angular velocity $\Omega_{\rm p}$, the two centers have fixed positions $C_1(x,y) = \left(x_1,0\right)$ and $C_2(x,y) = \left(x_2,0\right)$, respectively. The total gravitational potential which is responsible for the motion of a star in the dynamical system of the binary galaxy is
\begin{equation}
\Phi_{\rm t}(x,y)=\Phi_1(x,y) + \Phi_2(x,y) + \Phi_{\rm rot}(x,y),
\label{pot}
\end{equation}
where
\[
\Phi_1(x,y) = \frac{- \ G M_1}{\sqrt{R_1^2 + c_1^2}},
\]
\[
\Phi_2(x,y) = \frac{- \ G M_2}{\sqrt{R_2^2 + c_2^2}},
\]
\begin{equation}
\Phi_{rot}(x,y) = - \frac{\Omega_p^2}{2}\left(x^2 + y^2 \right),
\end{equation}
and
\begin{equation}
R_1^2 = \left(x - x_1\right)^2 + y^2, \ \ \ R_2^2 = \left(x - x_2\right)^2 + y^2,
\end{equation}
with
\begin{equation}
x_1 = -\frac{M_2}{M_t} \ d, \ \ \ x_2 = R - \frac{M_2}{M_t} \ d = d + x_1.
\end{equation}

In our study we use a system of galactic units where the unit of length is 20 kpc, the unit of mass is $1.8 \times 10^{11} {\rm M}_\odot$ and the unit of time is $0.99 \times 10^8$ yr. The velocity unit is 197 km/s, while $G$ is equal to unity ($G$ = 1). In these units, we use the values: $M_1 = M_2 = 1$, $c_1 = c_2 = 0.2$ and $d = 2$. The values of these quantities remain constant throughout securing also positive mass density everywhere and free of singularities. The fact that the two galaxies are sufficiently apart from each other $(d$ = 40 kpc) allow us to assume that the tidal phenomena are very small and therefore negligible.

The two galaxies move around their common mass center of the system with angular velocities $\Omega_{p1}$ and $\Omega_{p2}$, given by
\[
\Omega_{\rm p1} = \sqrt{\frac{1}{x_1}\left(\frac{- \ d \Phi_2(R)}{dR}\right)_{R = d}},
\]
\begin{equation}
\Omega_{\rm p2} = \sqrt{\frac{1}{x_2}\left(\frac{d \Phi_1(R)}{dR}\right)_{R = d}}.
\end{equation}
Here it should be pointed out that the two angular velocities are equal $(\Omega_{\rm p1} = \Omega_{\rm p2}$ = 0.496282514512041) due to the fact that the two galaxies are identical having the same mass and equal core radii. Nevertheless, as the two galaxies are not perfect mass points $(c_1, c_2 \neq 0)$ the two angular velocities do not exactly coincide with the value derived by Kepler's third law $(\Omega_{\rm p} = 1/2)$. The deviation $|\Omega_{\rm p} - \Omega_{\rm p1}| = 0.0037$ however, is very small, almost negligible, and therefore for simplicity we can still compute the angular velocity through Kepler's third law. In the case where the two galaxies are not identical this assumption is not valid.

\begin{figure}[!tH]
\includegraphics[width=\hsize]{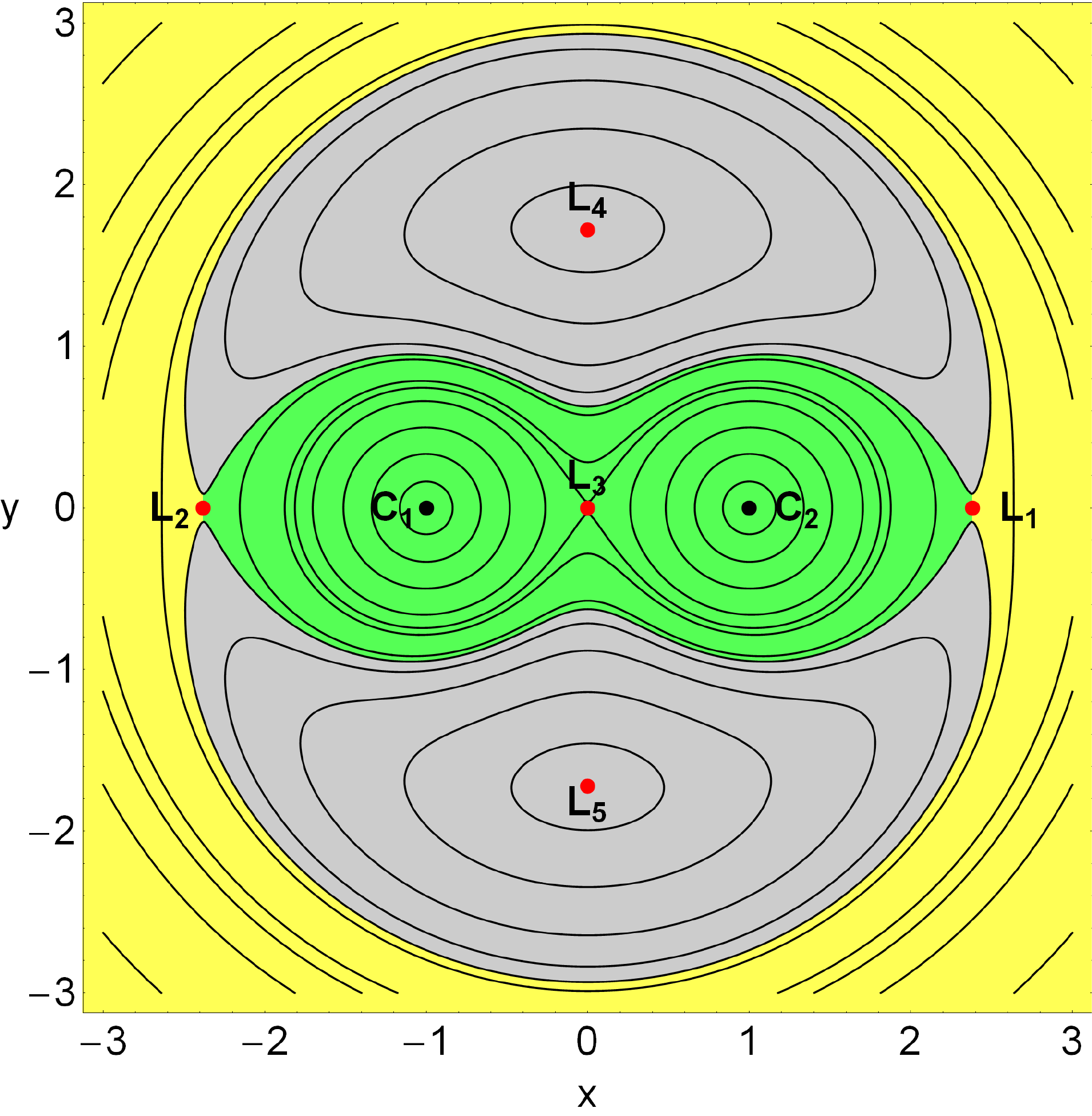}
\caption{The isolines contours of the constant total potential and the position of the five Lagrangian points. The escaping orbits leak out through exit channels 1 and 2 of the equipotential surface passing either $L_1$ or $L_2$, respectively. The interior region is indicated in green, the exterior region is shown in amber color, while the forbidden regions of motion are marked with grey.}
\label{conts}
\end{figure}

This binary galaxy model has five equilibria called Lagrangian points at which
\begin{equation}
\frac{\partial \Phi_{\rm t}}{\partial x} = \frac{\partial \Phi_{\rm t}}{\partial y} = 0.
\label{lps}
\end{equation}
The isolines contours of constant total potential as well as the position of the five Lagrangian points $L_i, \ i = {1,5}$ are shown in Fig. \ref{conts}. Three of them, $L_1$, $L_2$, and $L_3$, known as the collinear points, are located in the $x$-axis. The central stationary point $L_3$ at $(x,y) = (0,0)$ is a local minimum of $\Phi_{\rm t}$. At the other four Lagrangian points it is possible for the test particle (star) to move in a circular orbit, while appearing to be stationary in the rotating frame. For this circular orbit, the centrifugal and the gravitational forces precisely balance. The stationary points $L_1$ and $L_2$ at $(x,y) = (\pm r_L,0) = (\pm 2.381038250079726,0)$ are saddle points, where $r_L$ is called Lagrangian radius. Let $L_1$ located at $x = +r_L$, while $L_2$ be at $x = -r_L$. The points $L_4$ and $L_5$ on the other hand, which are located at $(x,y) = (0, \pm 1.720465053408526)$ are local maxima of the total potential, enclosed by the banana-shaped isolines. The annulus bounded by the circles through $L_1$, $L_2$ and $L_4$, $L_5$ is known as the ``region of coroation" (see also \citep{BT08}).

The equations of motion in the rotating frame are described in vectorial form by
\begin{equation}
\ddot{{\vec{r}}} = - {\vec{\nabla}} \Phi_{\rm t} - 2\left({\vec{\Omega}} \times \dot{{\vec{r}}} \right),
\label{eqmot0}
\end{equation}
where $\vec{r} = (x,y,z)$ is the position vector, $\vec{\Omega} = (0,0,\Omega)$ is the constant rotation velocity vector around the vertical $z$-axis, while the term $- 2\left(\vec{\Omega} \times \dot{\vec{r}} \right)$ represents the Coriolis force. Decomposing Eq. (\ref{eqmot0}) into rectangular Cartesian coordinates $(x,y)$, we obtain
\begin{eqnarray}
\ddot{x} = - \frac{\partial \Phi_{\rm t}}{\partial x} + 2\Omega\dot{y}, \nonumber \\
\ddot{y} = - \frac{\partial \Phi_{\rm t}}{\partial y} - 2\Omega\dot{x},
\label{eqmot}
\end{eqnarray}
where the dot indicates derivative with respect to the time. The standard variational method is used for the variational equations needed for the calculation of the chaos indicator (SALI).

Consequently, the corresponding Hamiltonian (known also as the Jacobian) to the total potential given in Eq. (\ref{pot}) reads
\begin{equation}
H_{\rm J}(x,y,\dot{x},\dot{y}) = \frac{1}{2} \left(\dot{x}^2 + \dot{y}^2 \right) + \Phi_{\rm t}(x,y) = E,
\label{ham}
\end{equation}
where $\dot{x}$ and $\dot{y}$ are the momenta per unit mass, conjugate to $x$ and $y$, respectively, while $E$ is the numerical value of the Jacobian, which is conserved since it is an isolating integral of motion. Thus, an orbit with a given value for it's energy integral is restricted in its motion to regions in which $E \leq \Phi_{\rm t}$, while all other regions are forbidden to the star. The numerical value of the total potential at the two Lagrangian points $L_1$ and $L_2$, that is $\Phi_{\rm t}(r_L,0)$ and $\Phi_{\rm t}(-r_L,0)$, respectively yields to a critical Jacobi constant $C_L = -1.720536218256113$, which can be used to define a dimensionless energy parameter as
\begin{equation}
\widehat{C} = \frac{C_L - E}{C_L},
\label{chat}
\end{equation}
where $E$ is some other value of the Jacobian. The dimensionless energy parameter $\widehat{C}$ makes the reference to energy levels more convenient. For $E = C_L$ the equipotential surface encloses the critical area\footnote{In three dimensions the last closed equipotential surface of the total potential passing through the Lagrangian points $L_1$ and $L_2$ encloses a critical volume.}, while for a Jacobian value $E > C_L$, or in other words when $\widehat{C} > 0$, the equipotential surface is open and consequently stars can escape from the system. In Fig. \ref{conts} we observe the two openings (exit channels) near $L_1$ and $L_2$ through which the stars can leak out. In fact, we may say that these two exits act as hoses connecting the interior region (green color) of the binary system where $-r_L < x < r_L$ with the ``outside world" of the exterior region (amber color). Exit channel 1 at the positive $x$-direction corresponds to escape through Lagrangian point $L_1$, while exit channel 2 at the negative $x$-direction indicates escape through $L_2$. The forbidden regions of motion within the banana-shaped isolines around $L_4$ and $L_5$ points are shown in Fig. \ref{conts} with gray. It is evident, that for $E < C_L$ the two forbidden regions merge together thus escape is impossible since the interior region cannon communicate with the exterior area.

In dynamical systems with escapes an issue of great importance is the determination of the position as well as the time at which an orbit escapes. An open equipotential surface consists of two branches forming channels through which an orbit can escape to infinity (see Fig. \ref{conts}). It was proved \citep{C79} that in Hamiltonian systems there is a highly unstable periodic orbit at every opening close to the line of maximum potential which is called a Lyapunov orbit. In fact, there is a family of unstable Lyapunov orbits around each collinear Lagrangian point \citep{M58}. Such an orbit reaches the Zero Velocity Curve\footnote{The boundaries of the accessible regions of motion are the called Zero Velocity Curves, since they are the locus in the $(x,y)$ space where kinetic energy vanishes.} (ZVC), on both sides of the opening and returns along the same path thus, connecting two opposite branches of the ZVC. Lyapunov orbits are very important for the escapes from the system, since if an orbit intersects any one of these orbits with velocity pointing outwards moves always outwards and eventually escapes from the system without any further intersections with the surface of section (e.g., \citet{C90}). When $E = C_L$ the Lagrangian points exist precisely but when $E > C_L$ an unstable Lyapunov periodic orbit is located close to each of these two points (e.g., \citet{H69}). These Lagrangian points $L_1$ and $L_2$ are saddle points of the total potential, so when $E > C_L$, a star must pass close enough to one of these points in order to escape. Thus, in our binary galaxy system the escape criterion is purely geometric. In particular, escapers are defined to be those stars moving in orbits beyond the Lagrangian radius $(r_L)$ thus passing one of the two Lagrangian points ($L_1$ or $L_2$) and intersecting one of the two unstable Lyapunov orbits with velocity pointing outwards. Here we must emphasize that orbits with initial conditions outside the interior region, or in other words outside $L_1$ or $L_2$, does not necessarily escape immediately from the galaxy. Thus the initial position itself does not furnish a sufficient condition for escape, since the escape criterion is a combination of the coordinates and the velocity of stars.

Undoubtedly, a binary system of two interacting galaxies is a very complex stellar entity and therefore we need to assume some necessary simplifications and assumptions in order to be able to study mathematically the orbital behavior of such a complicated stellar system. For this purpose, our total gravitational model is intentionally simple and contrived, in order to give us the ability to study all the different aspects regarding the kinematics and dynamics of the model. Nevertheless, contrived models can surely provide an insight into more realistic stellar systems, which unfortunately are very difficult to be explored, if we take into account all the astrophysical aspects (i.e., gas, spirals, mergers, etc). Self-consistent models on the other hand, are mainly used when conducting $N$-body simulations. However, this application is entirely out of the scope of the present paper. Once more, we have to point out that the mathematical simplicity of our model is necessary, otherwise it would be extremely difficult, or even impossible, to apply the detailed numerical calculations presented in this study. Similar gravitational binary models with the same limitations and assumptions were used successfully several times in the past in order to investigate the orbital structure in much more complicated galactic systems (e.g., \citet{CI09,CP09,CZ09,Z12b,Z13}).

\section{Computational methods}
\label{cometh}

For exploring the escape dynamics of stars in the binary galaxy model, we need to define samples of initial conditions of orbits whose properties (bounded or escaping motion) will be identified. For this purpose we define for each value of the total orbital energy (all tested energy levels are above the escape energy), dense uniform grids of $1024 \times 1024$ initial conditions regularly distributed in the area allowed by the value of the energy. Following a typical approach, all orbits are launched with initial conditions inside the Lagrangian radius $(x_0^2 + y_0^2 \leq r_L^2)$, or in other words the interior green region shown in Fig. \ref{conts} which is the scattering region in our case. Our investigation takes place both in the configuration $(x,y)$ and in the phase $(x,\dot{x})$ space for a better understanding of the escape process. Furthermore, the grids of initial conditions of orbits whose properties will be examined are defined as follows: For the configuration $(x,y)$ space we consider orbits with initial conditions $(x_0, y_0)$ with $\dot{x_0} = 0$, while the initial value of $\dot{y_0}$ is always obtained from the energy integral  of Eq. (\ref{ham}) as $\dot{y_0} = \dot{y}(x_0,y_0,\dot{x_0},E) > 0$. Similarly, for the phase $(x,\dot{x})$ space we consider orbits with initial conditions $(x_0, \dot{x_0})$ with $y_0 = 0$, while this time the initial value of $\dot{y_0}$ is obtained from the Jacobi integral of Eq. (\ref{ham}).

The equations of motion as well as the variational equations for the initial conditions of all orbits were integrated using a double precision Bulirsch-Stoer \verb!FORTRAN 77! algorithm (e.g., \citep{PTVF92}) with a small time step of order of $10^{-2}$, which is sufficient enough for the desired accuracy of our computations. Here we should emphasize, that our previous numerical experience suggests that the Bulirsch-Stoer integrator is both faster and more accurate than a double precision Runge-Kutta-Fehlberg algorithm of order 7 with Cash-Karp coefficients (e.g., \citep{DMCG12}). Throughout all our computations, the Jacobian energy integral (Eq. (\ref{ham})) was conserved better than one part in $10^{-11}$, although for most orbits it was better than one part in $10^{-12}$. For the numerical integration of the orbits we needed about between 1 hour and 9 days of CPU time on a Pentium Dual-Core 2.2 GHz PC, depending on the escape rates of orbits in each case.

The configuration and the phase space are divided into the escaping and non-escaping (trapped) space. Usually, the vast majority of the trapped space is occupied by initial conditions of regular orbits forming stability islands where a third integral is present. In many systems however, trapped chaotic orbits have also been observed. Therefore, we decided to distinguish between regular and chaotic trapped motion. Over the years, several chaos indicators have been developed in order to determine the character of orbits (e.g., the Lyapunov Characteristic Exponent or LCE \citep{BGGS80}; the Fast Lyapunov Indicator or FLI \citep{FGL97}; the Generalized ALignment Index or GALI \citep{SBA07}; the Relative Lyapunov Indicator or RLI \citep{SESF04}; the Orthogonal Fast Lyapunov Indicator or OFLI \citep{FLFF02}; and the Mean Exponential Growth factor of Nearby Orbits or MENGO \citep{CGS03}). In our case, we chose to use the Smaller ALingment Index (SALI) method. The SALI \citep{S01} has been proved a very fast, reliable and effective tool, which is defined as
\begin{equation}
\rm SALI(t) \equiv min(d_-, d_+),
\label{sali}
\end{equation}
where $d_- \equiv \| {\vec{w_1}}(t) - {\vec{w_2}}(t) \|$ and $d_+ \equiv \| {\vec{w_1}}(t) + {\vec{w_2}}(t) \|$ are the alignments indices, while ${\vec{w_1}}(t)$ and ${\vec{w_2}}(t)$, are two deviation vectors which initially point in two random directions. For distinguishing between ordered and chaotic motion, all we have to do is to compute the SALI along time interval $t_{max}$ of numerical integration. In particular, we track simultaneously the time-evolution of the main orbit itself as well as the two deviation vectors ${\vec{w_1}}(t)$ and ${\vec{w_2}}(t)$ in order to compute the SALI.

The time-evolution of SALI strongly depends on the nature of the computed orbit since when an orbit is regular the SALI exhibits small fluctuations around non zero values, while on the other hand, in the case of chaotic orbits the SALI after a small transient period it tends exponentially to zero approaching the limit of the accuracy of the computer $(10^{-16})$. Therefore, the particular time-evolution of the SALI allow us to distinguish fast and safely between regular and chaotic motion. Nevertheless, we have to define a specific numerical threshold value for determining the transition from order to chaos. For this purpose we integrated a set of $10^4$ random initial conditions of orbits searching for a safe threshold value. Our numerical experiments suggest that a safe threshold value for the SALI is the value $10^{-8}$. Here we have to point out that the same threshold value was used in \citet{MA11}. Thus, in order to decide whether an orbit is regular or chaotic, one may follow the usual method according to which we check after a certain and predefined time interval of numerical integration, if the value of SALI has become less than the established threshold value. Therefore, if SALI $\leq 10^{-8}$ the orbit is chaotic, while if SALI $ > 10^{-8}$ the orbit is regular thus making the distinction between regular and chaotic motion clear and safe. When $10^{-6} < $ SALI $ < 10^{-8}$ we have the phenomenon of weak chaos where the corresponding orbits are irregular or not yet chaotic. For the computation of SALI we used the \verb!LP-VI! code \citep{CMD14}, a fully operational routine which efficiently computes a suite of many chaos indicators for dynamical systems in any number of dimensions.

In our computations, we set $10^4$ time units as a maximum time of numerical integration\footnote{In \citet{MVS04} the authors indicated that simulations in static potentials should be constrained to roughly 5 Gyr, because the potentials could evolve at larger time scales.}. The vast majority of orbits (regular and chaotic) however, need considerable less time to find one of the two exits in the limiting curve and eventually escape from the system (obviously, the numerical integration is effectively ended when an orbit passes through one of the escape channels and escapes). Nevertheless, we decided to use such a vast integration time just to be sure that all orbits have enough time in order to escape. Remember, that there are the so called ``sticky orbits" which behave as regular ones during long periods of time. Here we should clarify, that orbits which do not escape after a numerical integration of $10^4$ time units are considered as non-escaping or trapped. In fact, orbits with escape periods equal to many Hubble times are completely irrelevant to our investigation since they lack physical meaning.

\section{Numerical results}
\label{numres}

\begin{figure*}[!tH]
\centering
\resizebox{0.8\hsize}{!}{\includegraphics{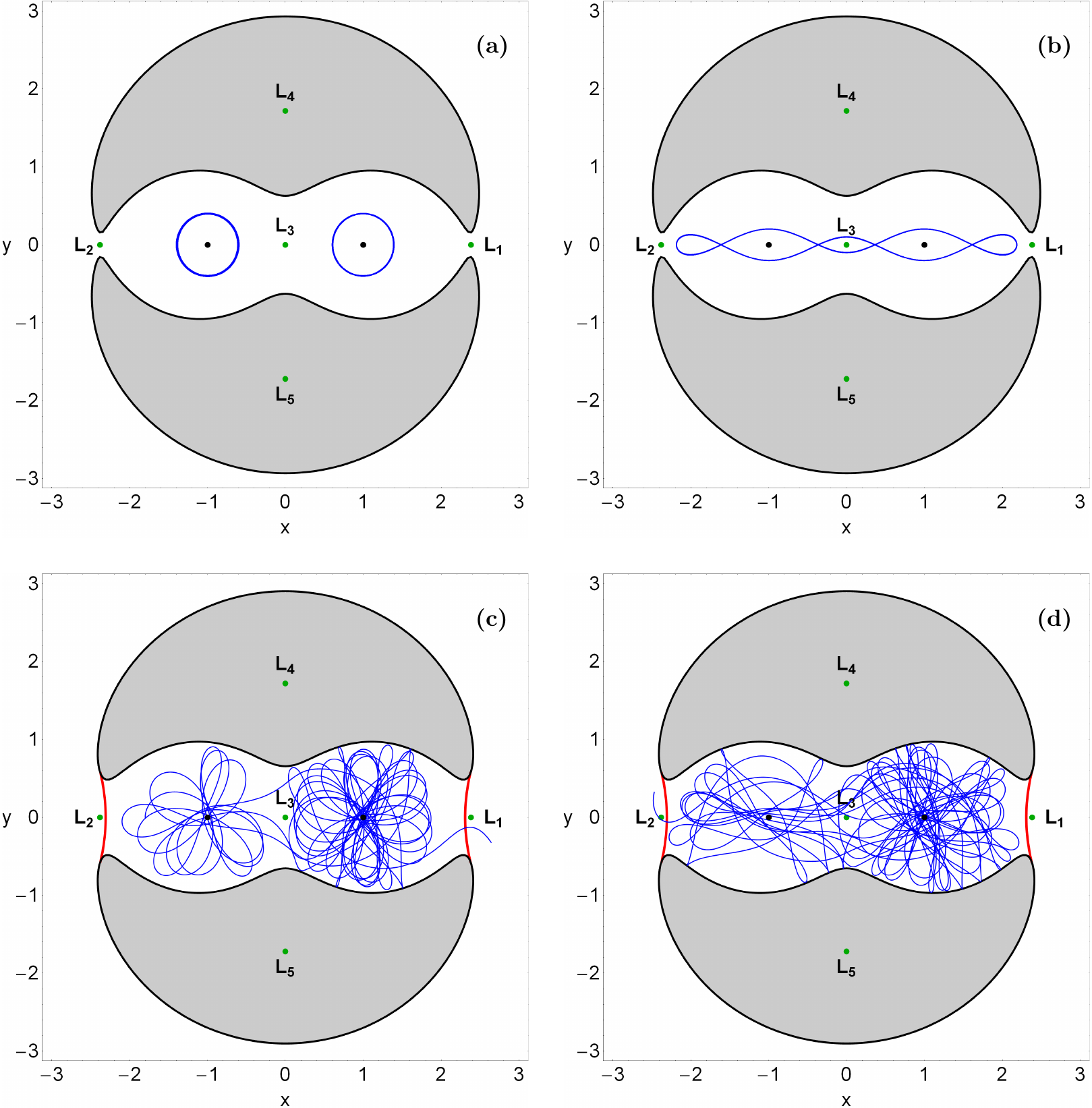}}
\caption{Characteristic examples of the main types of orbits in our binary galaxy model. (a-upper left): two 1:1 loop orbits circulating around one of the two galaxies, (b-upper right): a 1:5 resonant quasi-periodic box orbit circulating around both galaxies, (c-lower left): an orbit escaping through $L_1$, (d-lower right): an orbit escaping through $L_2$. The two unstable Lyapunov orbits around the Lagrangian points are visualized in red. More details are given in Table \ref{table1}.}
\label{orbs}
\end{figure*}

In this section we shall perform an orbit classification determining which orbits escape and which remain trapped, distinguishing simultaneously between regular and chaotic trapped motion\footnote{Generally, any dynamical method requires a sufficient time interval of numerical integration in order to distinguish safely between ordered and chaotic motion. Therefore, if the escape rate of an orbit is very low or even worse if the orbit escapes directly from the system then, any chaos indicator (the SALI in our case) will fail to work properly due to insufficient integration time. Hence, it is pointless to speak of regular or chaotic escaping orbits.}. Furthermore, two additional properties of the orbits will be examined: (i) the channels or exits through which the stars escape and (ii) the time-scale of the escapes (we shall also use the terms escape period or escape rates). We will investigate these dynamical quantities for various values of the energy, always within the interval $\widehat{C} \in [0.001,0.1]$.

Our numerical calculations indicate that apart from the escaping orbits there is always a considerable amount of non-escaping orbits. In general terms, the majority of non-escaping regions corresponds to initial conditions of regular orbits, where a third integral of motion is present, restricting their accessible phase space and therefore hinders their escape. However, there are also chaotic orbits which do not escape within the predefined interval of $10^4$ time units and remain trapped for vast periods until they eventually escape to infinity. At this point, it should be emphasized and clarified that these trapped chaotic orbits cannot be considered, by no means, neither as sticky orbits nor as super sticky orbits with sticky periods larger than $10^4$ time units. In our case on the other hand, this type of orbits exhibit chaoticity very quickly as it takes no more than about 200 time units for the SALI to cross the threshold value (SALI $\ll 10^{-8}$), thus identifying their chaotic character. Therefore, we decided to classify the initial conditions of orbits in both the configuration and phase space into four main categories: (i) orbits that escape through $L_1$, (ii) orbits that escape through $L_2$, (iii) non-escaping regular orbits and (iv) trapped chaotic orbits.

\begin{table}
\centering
\setlength{\tabcolsep}{2pt}
\begin{center}
   \centering
   \caption{Type, category, initial conditions, value of the energy integration and escape time of the orbits shown in Fig. \ref{orbs}(a-d). For all orbits we have $y_0 = \dot{x_0} = 0$, while the initial value of $\dot{y}$ is obtained from the Jacobi integral (\ref{ham}).}
   \label{table1}
   \begin{tabular}{@{}lcccccc}
      \hline
      Figure & Type & $x_0$ & $\widehat{C}$ & $t_{\rm int}$ & $t_{\rm esc}$ & Category \\
      \hline
      \ref{orbs}a & 1:1 loop & -1.400 & 0.001 & 100 &  -  & Non-escaping regular \\
      \ref{orbs}a & 1:1 loop &  0.600 & 0.001 & 100 &  -  & Non-escaping regular \\
      \ref{orbs}b & 1:5 box  & -2.185 & 0.001 & 100 &  -  & Non-escaping regular \\
      \ref{orbs}c & chaotic  & -1.800 & 0.010 & 116 & 111 & Escaping through $L_1$ \\
      \ref{orbs}d & chaotic  &  1.110 & 0.010 & 177 & 172 & Escaping through $L_2$ \\
      \hline
   \end{tabular}
\end{center}
\end{table}

Additional numerical computations reveal that the non-escaping regular orbits are mainly 1:1 loop orbits for which a third integral applies, while other types of secondary resonant orbits are also present. In Fig. \ref{orbs}a we present two 1:1 thin loop orbits, while in Fig. \ref{orbs}b a typical example of a secondary 1:5 resonant quasi-periodic orbit is shown. We observe that the 1:1 loop orbits circulate only around one of the two galaxies, while the 1:5 resonance orbit move around both galaxies. The $n:m$ notation we use for the regular orbits is according to \citet{CA98} and \citet{ZC13}, where the ratio of those integers corresponds to the ratio of the main frequencies of the orbit, where main frequency is the frequency of greatest amplitude in each coordinate. Main amplitudes, when having a rational ratio, define the resonances of an orbit. Finally in Figs. \ref{orbs}(c-d) we observe two orbits escaping through $L_1$ (exit channel 1) and $L_2$ (exit channel 2), respectively. The two regular orbits shown in Figs. \ref{orbs}(a-b) were computed until $t = 100$ time units, while on the other hand, the escaping orbits presented in Figs. \ref{orbs}(c-d) were calculated for 5 time units more than the corresponding escape period in order to visualize better the escape trail. For all orbits we have $y_0 = 0$, while the initial value of $\dot{y}$ is obtained from the Jacobi integral (\ref{ham}). The two unstable Lyapunov orbits around $L_1$ and $L_2$ are shown in Fig. \ref{orbs}(c-d) in red. In Table \ref{table1} we provide the type, the exact initial conditions and the value of the energy for all the depicted orbits.

\subsection{Structure of the configuration $(x,y)$ space}
\label{pp1}

\begin{figure*}[!tH]
\centering
\resizebox{0.7\hsize}{!}{\includegraphics{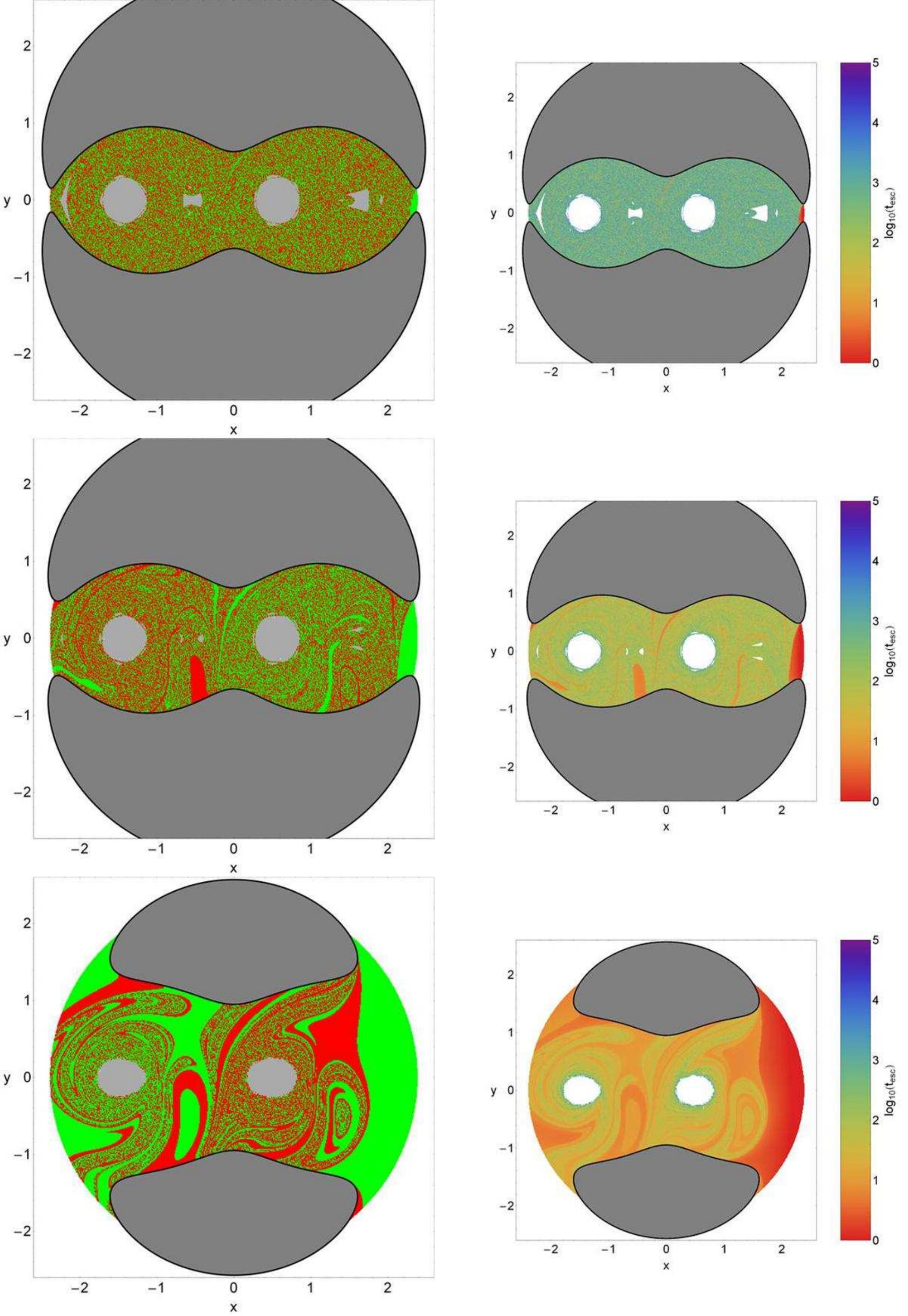}}
\caption{Left column: Orbital structure of the configuration $(x,y)$ space. Top row: $\widehat{C} = 0.001$; Middle row: $\widehat{C} = 0.01$; Bottom row: $\widehat{C} = 0.1$. The green regions correspond to initial conditions of orbits where the stars escape through $L_1$, red regions denote initial conditions where the stars escape through $L_2$, light gray areas represent stability islands of regular non-escaping orbits, initial conditions of trapped chaotic orbits are marked in white, while the regions of forbidden motion are shown in dark gray. Right column: Distribution of the corresponding escape times $t_{\rm esc}$ of the orbits on the configuration space. The darker the color, the larger the escape time. The scale on the color-bar is logarithmic. Initial conditions of non-escaping regular orbits and trapped chaotic orbits are shown in white. We decided to put together these two columns, so the reader can easily observe and correlate the escape basins with the corresponding escape times.}
\label{grd1}
\end{figure*}

Our exploration begins in the configuration space and in the left column of Fig. \ref{grd1} we present the orbital structure of the $(x,y)$ plane for three values of the energy, where the initial conditions of the orbits are classified into four categories by using different colors. Specifically, light gray color corresponds to regular non-escaping orbits, white color corresponds to trapped chaotic orbits, green color corresponds to orbits escaping through channel 1, while the initial conditions of orbits escaping through exit channel 2 are marked with red color. For $\widehat{C} = 0.001$, that is an energy level just above the critical escape energy $C_L$, we see that the vast majority of initial conditions corresponds to escaping orbits however, five stability islands of non-escaping regular orbits are present denoting ordered motion around the two galaxies. We also see that the entire interior region is completely fractal\footnote{It should be emphasized that the fractality of the structures was not measured by computing the corresponding fractal dimension. When we state that an area is fractal we mean that it has a fractal-like geometry.} which means that there is a highly dependence of the escape mechanism on the particular initial conditions of the orbits. In other words, a minor change in the initial conditions has as a result the star to escape through the opposite exit channel, which is of course, a classical indication of chaotic motion. With a much closer look at the $(x,y)$ plane we can identify some additional tiny stability islands which are deeply buried in the escape domain. As we proceed to higher energy levels the fractal area reduces and several basins of escape start to emerge. By the term basin of escape, we refer to a local set of initial conditions that corresponds to a certain escape channel. Indeed for $\widehat{C} = 0.01$ we observe the presence of escape basins which mainly have the shape of thin elongated bands. Moreover, the extent of the stability islands seems to be unaffected by the increase in the orbital energy. With increasing energy the interior region in the configuration spaces becomes less and less fractal and broad, well-defined basins of escape dominate when $\widehat{C} = 0.1$. The fractal regions on the other hand, are confined mainly near the boundaries between the escape basins or in the vicinity of the stability islands. Furthermore, it is seen that for $\widehat{C} = 0.1$ only two stability islands are present at the left part of the centers of the galaxies, while all smaller stability regions have disappeared. It should be emphasized that our classification shows that in all three energy levels trapped chaotic motion is almost negligible as the initial conditions of such orbits appear only as lonely points around the boundaries of the stability islands. Here we must point out two interesting phenomena that take place with increasing energy: (i) the regions of forbidden motion are significantly confined, (ii) the two escape channels become more and more wide.

\begin{figure}[!tH]
\includegraphics[width=\hsize]{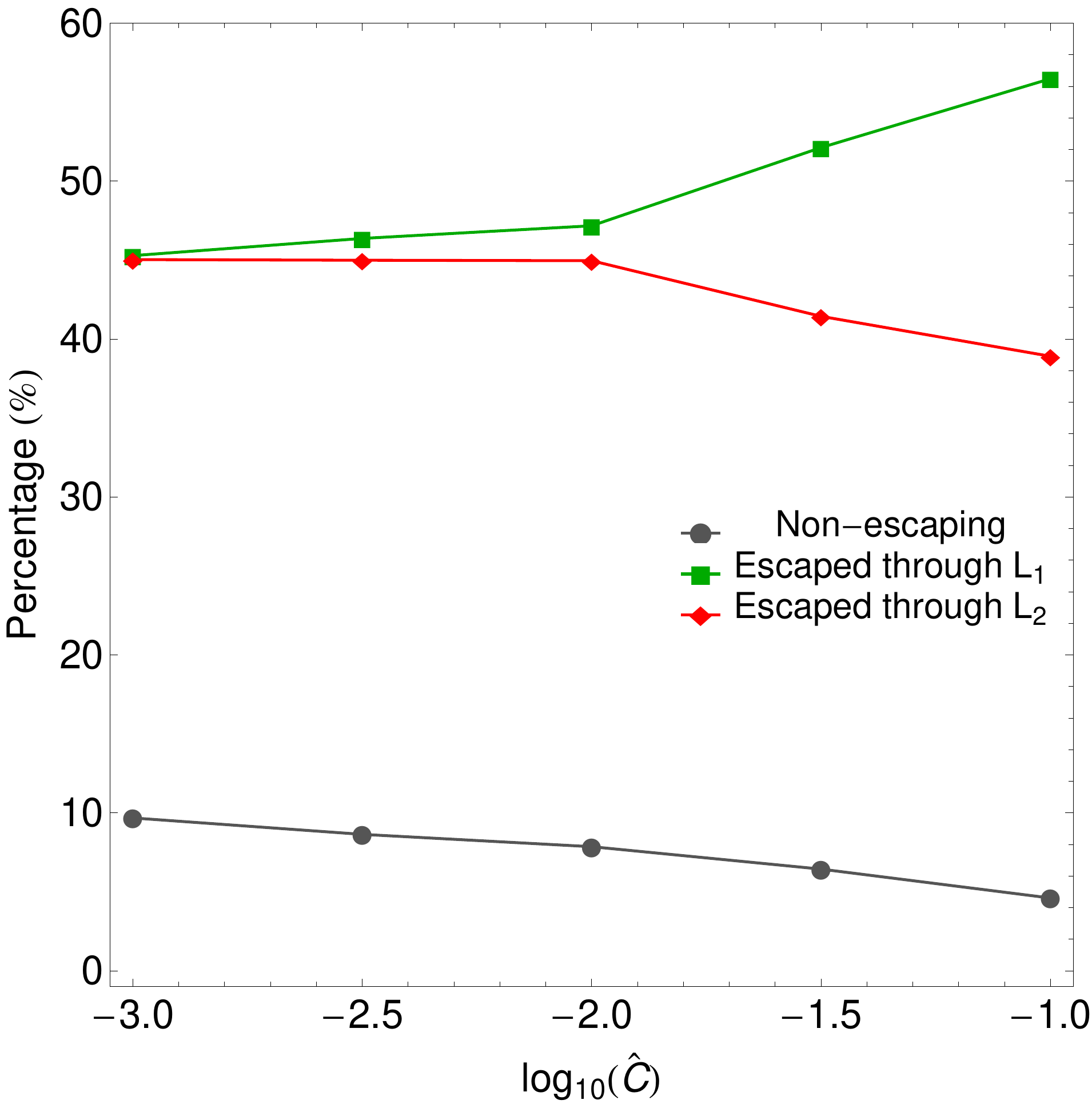}
\caption{Evolution of the percentages of escaping and non-escaping (regular plus chaotic) orbits on the configuration $(x,y)$ space when varying the dimensionless energy parameter $\widehat{C}$.}
\label{percs1}
\end{figure}

In the right column of Fig. \ref{grd1} we show how the escape times $t_{\rm esc}$ of orbits are distributed on the configuration $(x,y)$ space. The escape time $t_{\rm esc}$ is defined as the time when a star crosses one of the Lyapunov orbits with velocity pointing outwards. Light reddish colors correspond to fast escaping orbits with short escape periods, dark blue/purpe colors indicate large escape rates, while white color denote both non-escaping regular and trapped chaotic orbits. The scale on the color bar is logarithmic. It is evident, that orbits with initial conditions close to the boundaries of the stability islands need significant amount of time in order to escape from the system of two galaxies, while on the other hand, inside the basins of escape where there is no dependence on the initial conditions whatsoever, we measured the shortest escape rates of the orbits. We observe that for $\widehat{C} = 0.001$ the escape periods of orbits with initial conditions in the fractal region are huge corresponding to tens of thousands of time units. As the value of the total orbital energy increases however, the escape times of orbits reduce significantly. In fact for $\widehat{C} = 0.01$ and $\widehat{C} = 0.1$ the basins of escape can be distinguished in the grids of Fig. \ref{grd1}, being the regions with reddish colors indicating extremely fast escaping orbits. Our numerical calculations indicate that orbits with initial conditions inside the basins have significantly small escape periods of less than 10 time units.

In \citet{MA11} the authors showed how the percentages of regular and chaotic orbits in the phase space evolve as a function of the total energy and the main parameters of the system. In the same vein, in our case it would be very interesting to monitor how the percentages of the escaping and non-escaping orbits (instead of regular vs. chaotic) on the configuration $(x,y)$ space evolve when the dimensionless energy parameter $\widehat{C}$ varies. Such a diagram is presented in Fig. \ref{percs1}. This diagram is very helpful because as it quantifies the results shown in Fig. \ref{grd1}. At this point we must clarify that we decided to merge the percentages of non-escaping regular and trapped chaotic orbits together because our computations indicate that always the rate of trapped chaotic orbits is extremely small (less than 0.01\%) and therefore it does not contribute to the overall orbital structure of the dynamical system. One may observe, that when $\widehat{C} = 0.001$, that is just above the critical escape energy $C_L$, escaping orbits through $L_1$ and $L_2$ share about 90\% of the available configuration space, so the two exit channels can be considered equiprobable since the configuration space is highly fractal. As the value of the energy increases however, the percentages of escaping orbits through exit channels 1 and 2 start to diverge, especially for $\widehat{C} > 0.01$. In particular, the amount of orbits that escape through $L_1$ exhibits a linear increase, while on the other hand, the rate of escaping orbits through the opposite exit channel displays a linear decrease. At the highest energy level studied $(\widehat{C} = 0.1)$ escaping orbits through $L_1$ is the most populated family occupying about 56\% of the configuration plane, while only about 38\% of the same plane corresponds to initial conditions of orbits that escape through $L_2$. Furthermore, the percentage of non-escaping (regular plus chaotic) orbits is much less affected by the shifting of the energy displaying a minor decrease form 10\% to about 5\%. Thus taking into account all the above-mentioned analysis we may conclude that at low energy levels where the fractility of the configuration space is maximum the stars do not show any particular preference regarding the escape channel, while on the contrary, at high enough energy levels where basins of escape dominate it seems that exit channel 1 is more preferable\footnote{Even though both ext channels are geometrically symmetric the rotation of the two galaxies induce an asymmetry regarding the orbital structure and the escape basins. This is the main reason why for high enough values of the energy most orbits escape through channel 1.}.

\subsection{Structure of the phase $(x,\dot{x})$ space}
\label{pp2}

\begin{figure*}[!tH]
\centering
\resizebox{0.8\hsize}{!}{\includegraphics{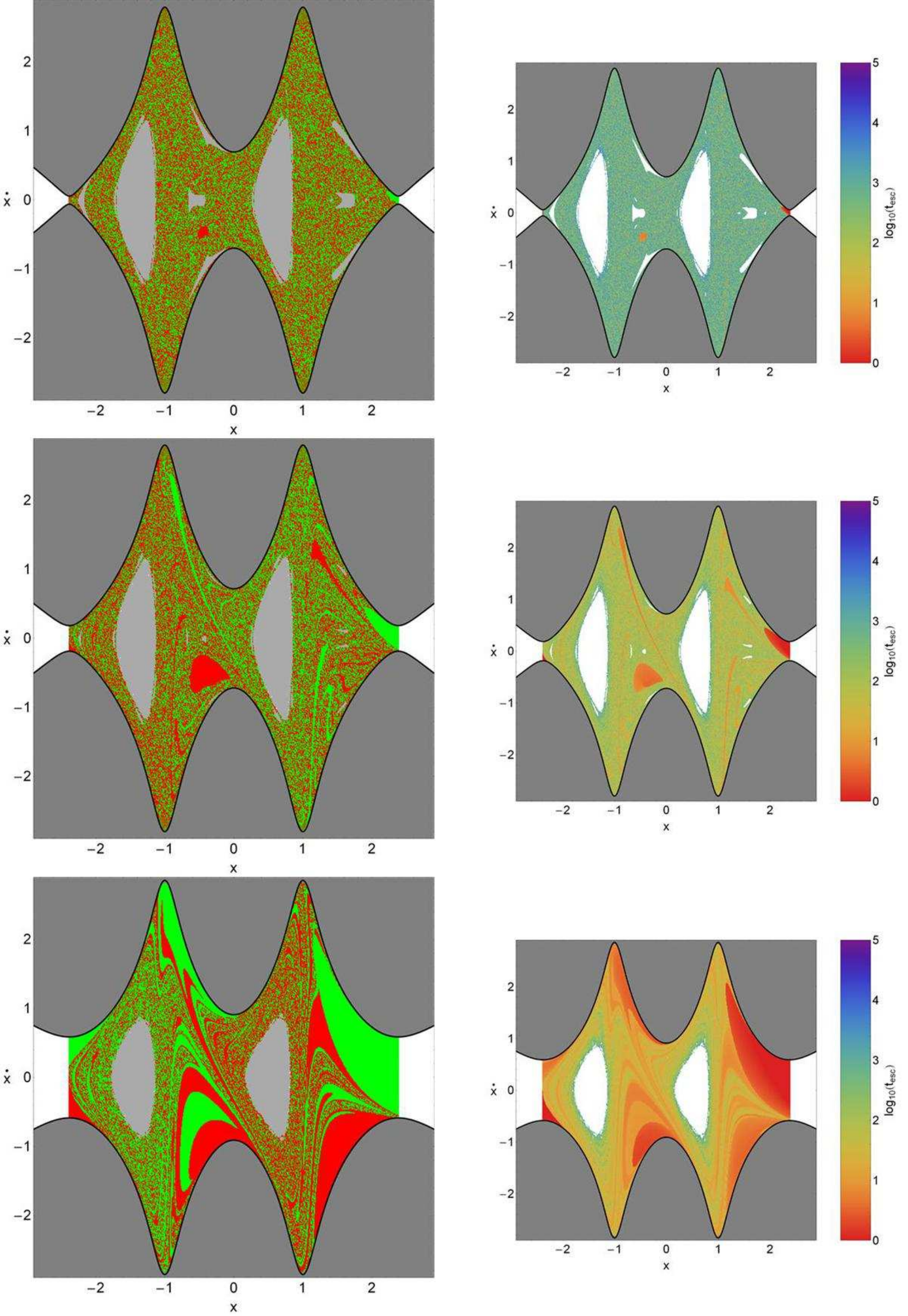}}
\caption{Left column: Orbital structure of the phase $(x,\dot{x})$ space. Top row: $\widehat{C} = 0.001$; Middle row: $\widehat{C} = 0.01$; Bottom row: $\widehat{C} = 0.1$. Right column: Distribution of the corresponding escape times $t_{\rm esc}$ of the orbits on the phase space. The color codes are the same as in Fig. \ref{grd1}.}
\label{grd2}
\end{figure*}

We continue our investigation in the phase $(x,\dot{x})$ space and we follow the same numerical approach as discussed previously. In Fig. \ref{grd2} we depict the orbital structure of the $(x,\dot{x})$ plane for three values of the energy, using different colors in order to distinguish between the four main types of orbits (non-escaping regular; trapped chaotic; escaping through $L_1$ and escaping through $L_2$). Just above the escape energy, that is when $\widehat{C} = 0.001$, we see that the structure of the phase plane is very similar to that of the corresponding configuration plane shown in Fig. \ref{grd1}. It is observed that about 82\% of the $(x,\dot{x})$ plane is occupied by initial conditions of escaping orbits. Moreover, the vast escape domain is highly fractal, while basins of escape, if any, are negligible. In all studied cases the areas of regular motion correspond mainly to retrograde orbits (i.e., when a star revolves around one galaxy in the opposite sense with respect to the motion of the galaxy itself), while there are also some smaller stability islands of prograde orbits. The area on the phase plane covered by escape basins grows drastically with increasing energy and at high enough energy levels the fractal regions are confined mainly at the boundaries of the stability islands of non-escaping orbits. In particular, the escape basins first appear at mediocre values of energy $(\widehat{C} = 0.01)$ inside the fractal, extended escape region having the shape of thin elongated bands however, as we proceed to higher energy levels they grow in size dominating in the phase space as well-formed broad domains. The Coriolis forces dictated by the rotation of the two spheroidal galaxies makes the phase planes to be asymmetric with respect to the $\dot{x}$-axis and this phenomenon is usually known as ``Coriolis asymmetry" (e.g., \citet{I80}). The distribution of the corresponding escape times $t_{\rm esc}$ of orbits on the phase space as a function of the energy parameter is shown in the right column Fig. \ref{grd2}. One may observe that the results are very similar to those presented earlier in Fig. \ref{grd1}, where we found that orbits with initial conditions inside the basins of escape have the smallest escape rates, while on the other hand, the longest escape times correspond to orbits with initial conditions either in the fractal regions of the plots, or near the boundaries of the islands of regular motion.

\begin{figure}[!tH]
\includegraphics[width=\hsize]{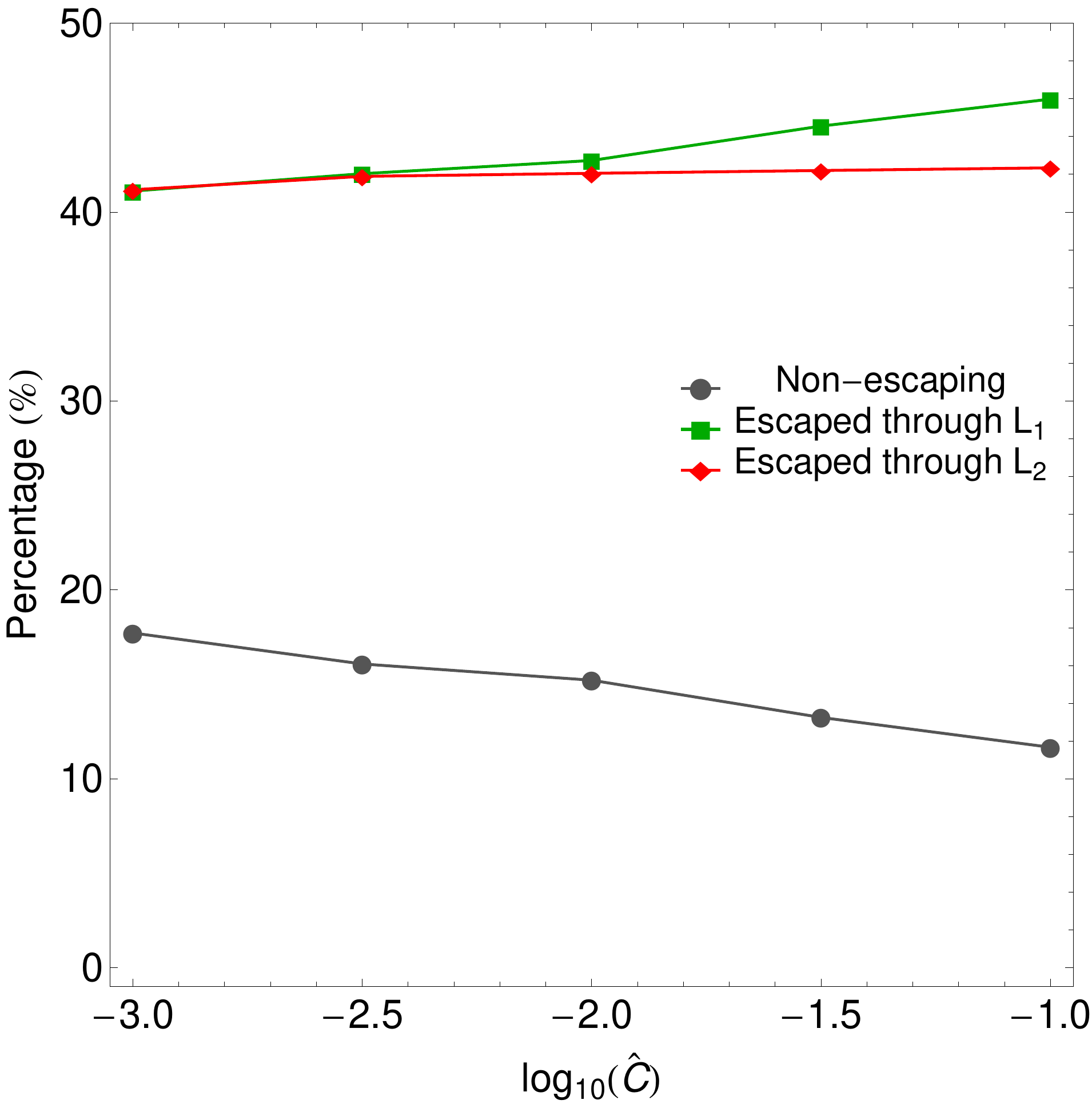}
\caption{Evolution of the percentages of escaping and non-escaping (regular plus chaotic) orbits on the phase $(x,\dot{x})$ space when varying the dimensionless energy parameter $\widehat{C}$.}
\label{percs2}
\end{figure}

The evolution of the percentages of the different types of orbits on the phase $(x,\dot{x})$ space as a function of the dimensionless energy parameter $\widehat{C}$ is presented in Fig. \ref{percs2}. Once more, we have to point out that the percentages of non-escaping ordered and trapped chaotic orbits are merged together in a single trend line because the percentage of trapped chaotic orbits is extremely small (less than 0.005\%) throughout. It is seen, that at the lowest energy level studied $(\widehat{C} = 0.001)$ escaping orbits through $L_1$ and $L_2$ share about 84\% of the phase plane, while only 16\% of the same plane corresponds to initial conditions of non-escaping (regular plus chaotic) orbits. Furthermore, we observe that for $0.001 < \widehat{C} < 0.01$ the rates of all types of orbits remain practically unperturbed by the shifting on the value of the orbital energy. For larger values of the energy $(\widehat{C} > 0.01)$ however, the percentages of escaping orbits exhibit a small divergence, while the portion of non-escaping orbits displays a minor decrease and for $\widehat{C} = 0.1$ they cover about 12\% of the phase plane. Specifically, the rate of orbits escaping through $L_1$ slightly increases reaching about 46\% at $\widehat{C} = 0.1$, while that of orbits escaping through $L_2$ continues the monotone behavior. It should be emphasized that the divergence of the percentages of escaping orbits observed in the phase space is much smaller than that found to exist in the configuration space. Therefore, one may reasonably deduce that in the phase space the increase in the value of the energy does not influence significantly the orbital content of the dynamical system, thus leaving the percentages of the orbits almost the same throughout. In addition, since that the rates of escaping orbits remain unperturbed within the energy range, we may say that the two exit channels can be considered almost equiprobable in the phase space.

\subsection{The $(x,\widehat{C})$ and the $(y,\widehat{C})$ spaces}
\label{over}

\begin{figure}[!tH]
\includegraphics[width=\hsize]{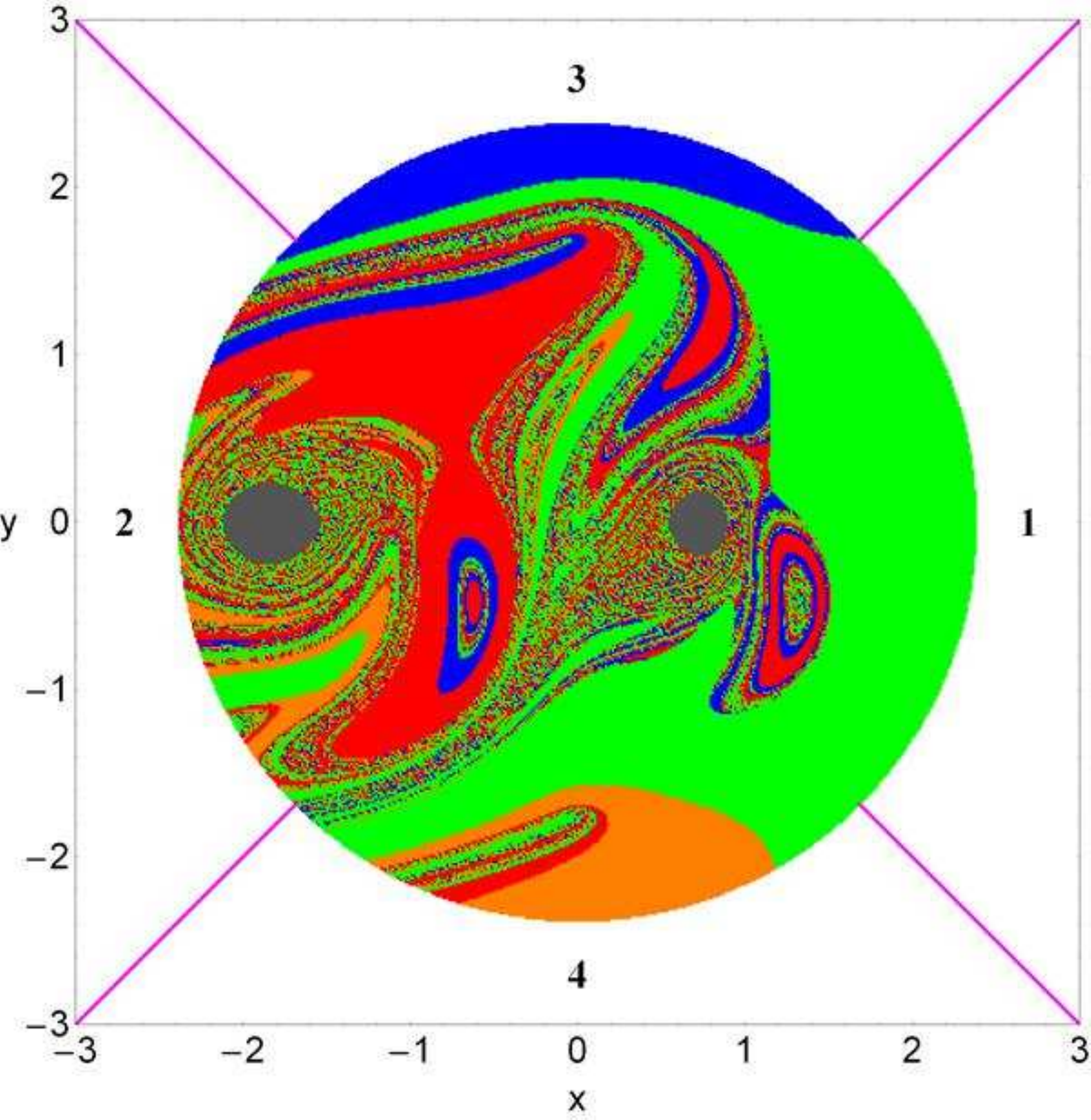}
\caption{Orbital structure of the configuration $(x,y)$ space for $E = E(L_4) = -1.37$. The basins of the four sectors are: sector 1 (green), sector 2 (red), sector 3 (blue), sector 4 (orange). Light gray areas represent stability islands of regular non-escaping orbits, while initial conditions of trapped chaotic orbits are marked in white. The magenta lines delimit the boundaries of the four sectors.}
\label{ll}
\end{figure}

\begin{figure*}[!tH]
\centering
\resizebox{0.9\hsize}{!}{\includegraphics{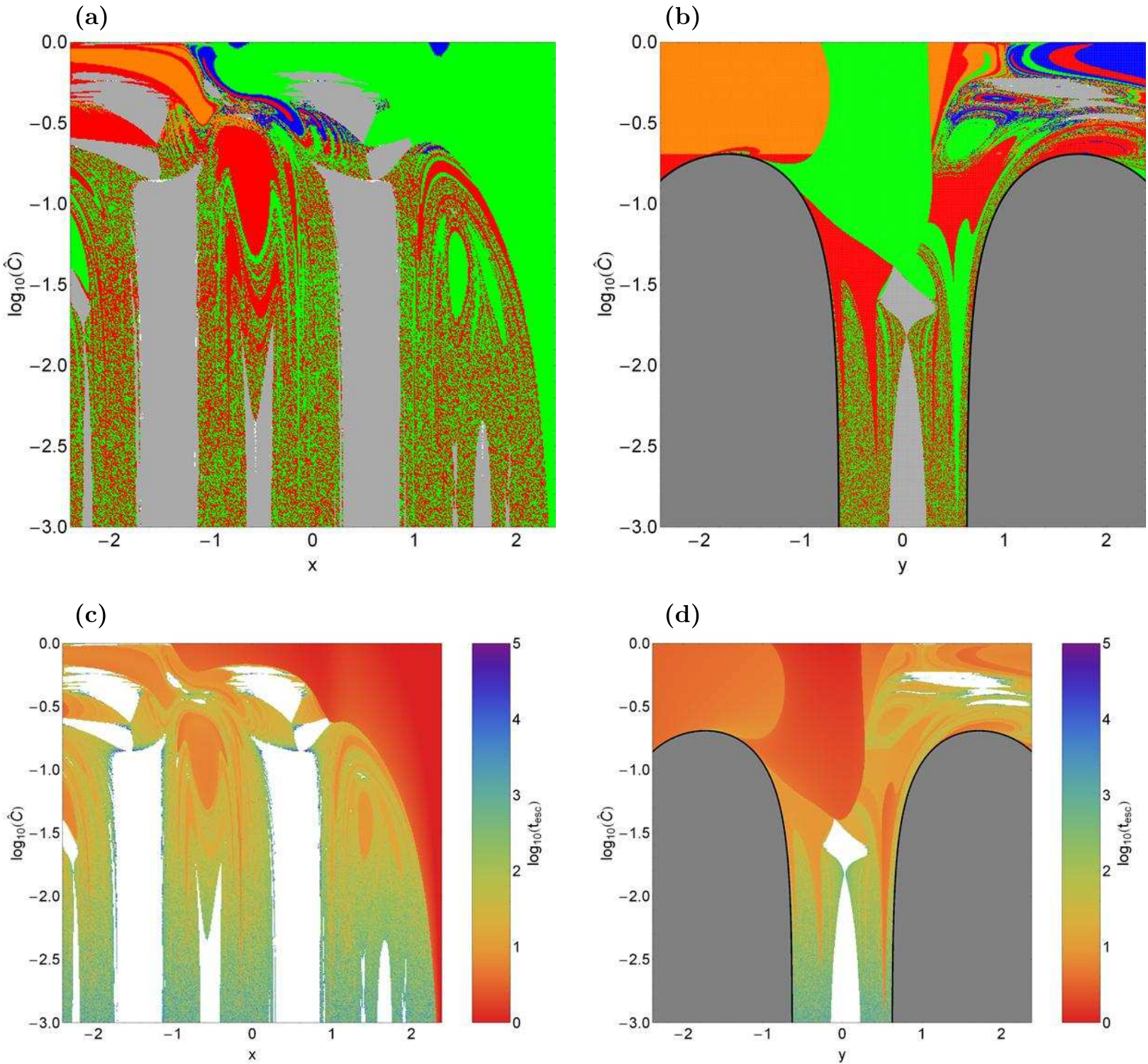}}
\caption{Orbital structure of the (a-upper left): $(x,\widehat{C})$-plane; and (b-upper right): $(y,\widehat{C})$-plane; (c-lower left and d-lower right): the distribution of the corresponding escape times of the orbits. The color codes are the exactly same as in Fig. \ref{grd1}.}
\label{xyc}
\end{figure*}

\begin{figure*}[!tH]
\centering
\resizebox{0.9\hsize}{!}{\includegraphics{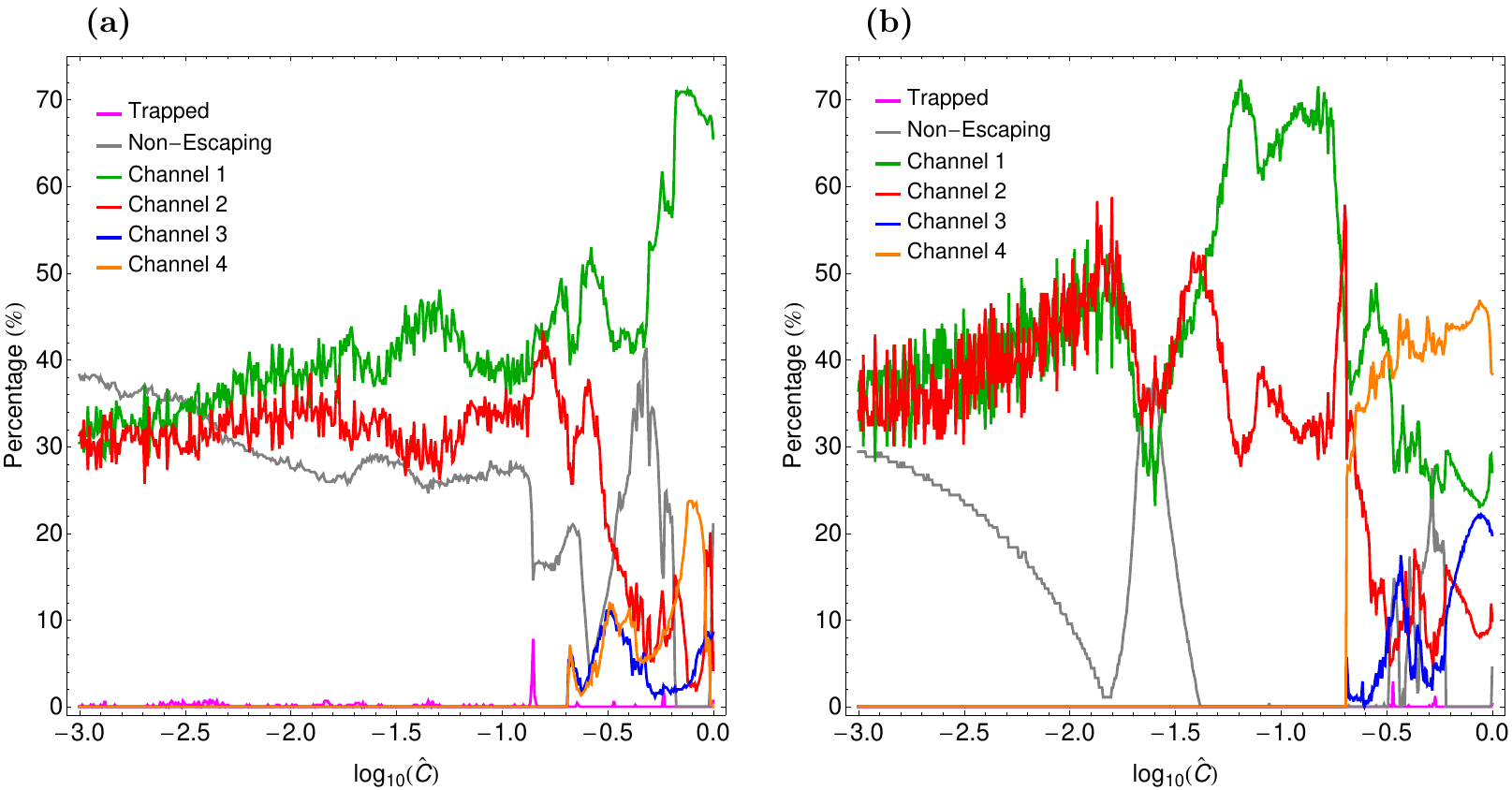}}
\caption{Evolution of the percentages of escaping, non-escaping regular and trapped chaotic orbits on the (a-left): $(x,\widehat{C})$-plane and (b-right): $(y,\widehat{C})$-plane as a function of the dimensionless energy parameter $\widehat{C}$.}
\label{percs3}
\end{figure*}

\begin{figure*}[!tH]
\centering
\resizebox{0.7\hsize}{!}{\includegraphics{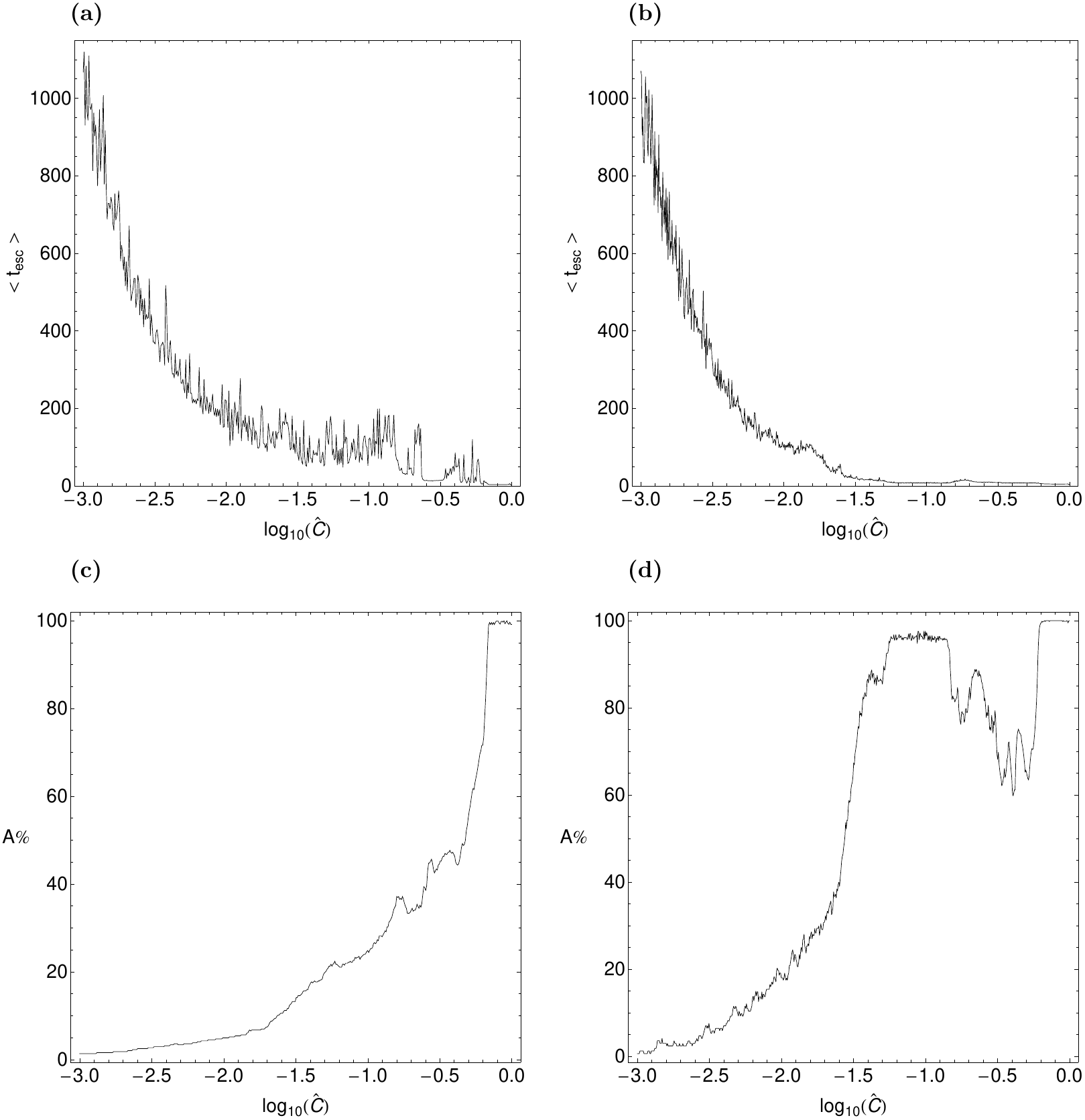}}
\caption{(a-b): The average escape time of orbits $< t_{\rm esc} >$ and (c-d): the percentage of the total area $(A\%)$, of the planes $(x,\widehat{C})$ and $(y,\widehat{C})$, respectively, covered by the escape basins as a function of the dimensionless energy parameter $\widehat{C}$.}
\label{stats}
\end{figure*}

The color-coded grids in the configuration $(x,y)$ as well as in the phase $(x,\dot{x})$ space provide sufficient information on the phase space mixing however, for only a fixed value of the Jacobi constant. H\'{e}non back in the late 60s \citep{H69}, introduced a new type of plane which can provide information not only about stability and chaotic regions but also about areas of trapped and escaping orbits using the section $y = \dot{x} = 0$, $\dot{y} > 0$ (see also \citet{BBS08}). In other words, all the orbits of the stars of the binary galaxy system are launched from the $x$-axis with $x = x_0$, parallel to the $y$-axis $(y = 0)$. Consequently, in contrast to the previously discussed types of planes, only orbits with pericenters on the $x$-axis are included and therefore, the value of the dimensionless energy parameter $\widehat{C}$ can be used as an ordinate. In this way, we can monitor how the energy influences the overall orbital structure of our dynamical system using a continuous spectrum of energy values rather than few discrete energy levels. We decided to explore the energy range when $\widehat{C} \in [0.001,1]$.

Here it should be pointed out that for $E > E(L_4) = -1.37$ the forbidden regions disappear and all the phase space is available for motion. Therefore for $E > -1.37$ we need to introduce new escape criteria. In particular, we divide the $(x,y)$ plane into four sectors according to a polar angle $\theta$ which starts counting from the positive part of the $x$-axis $(x > 0, y = 0)$ using the approach followed in \citet{dAT14}. So, considering that $\theta \in [0^{\circ},360^{\circ}]$, we have the first sector for $\theta \geq 315^{\circ}$ or $\theta < 45^{\circ}$, the second sector for $135^{\circ} \leq \theta < 225^{\circ}$, the third sector for $45^{\circ} \leq \theta < 135^{\circ}$ and the fourth sector for $255^{\circ} \leq \theta < 315^{\circ}$, respectively. We define the sectors in such a way so that sectors 1 and 2 to correspond to the previous escape channels 1 and 2, respectively. In Fig. \ref{ll} we present the structure of the configuration $(x,y)$ plane for $E = E(L_4) = -1.37$ where the initial conditions of the escaping orbits are colored according to the four sectors (the magenta lines denote the limits of each sector). It is seen that most of the orbits escape through sectors 1 and 2, while the basins corresponding to sectors 3 and 4 are smaller. So, despite the ZVC is not so restrictive when $E > E(L_4)$, escape occur mostly to the right and to the left of the scattering region. In particular, for $E = -1.37$, 46.68\% of the total integrated initial conditions belongs to the basin of the first sector, 27.16\% belongs to the basin of the second sector, 13.40\% belongs to the basin of the third sector, and just 10.93\% belongs to the basin of the fourth sector. The remaining 1.83\% of the solutions belong to the bounded basin.

In Fig. \ref{xyc}a we present the orbital structure of the $(x,\widehat{C})$-plane when $\widehat{C} \in [0.001,1]$, while in Fig. \ref{xyc}c the distribution of the corresponding escape times of orbits is depicted. We observe that for relatively low energy values $(0.001 < \widehat{C} < 0.01)$ the interior region is highly fractal, while basins of escape are located only at the outer parts of the $(x,\widehat{C})$-plane, that is outside $L_1$ and $L_2$ (or in other words in the exterior region). Furthermore, we can identify the presence of two main stability islands of prograde $(x_0 > 0)$ non-escaping regular orbits and three stability islands of retrograde $(x_0 < 0)$ regular motion. For larger values of energy $\widehat{C} > 0.01$ however, the structure of the $(x,\widehat{C})$-plane changes drastically and the most important differences are the following: (i) several basins of escape are formed inside the fractal escape region, (ii) two main stability islands are present. It should be pointed out that in the blow-ups of the diagram several additional extremely tiny islands of stability have been identified\footnote{An infinite number of regions of (stable) quasi-periodic (or small scale chaotic) motion is expected from classical chaos theory.}, (iii) at high enough energy levels $(\widehat{C} > 0.1)$ the basin of escape corresponding to exit 1 or sector 1 cover the vast majority of the grid, (iv) the extent of the main island of regular motion grows rapidly and for $\widehat{C} > 0.1$ the position of initial conditions of non-escaping regular orbits exceed the interior region $(x_0 < - r_L)$.

In order to obtain a more complete view on how the total orbital energy influences the nature of orbits in our binary galaxy model, we follow a similar numerical approach to that explained before but now all orbits of stars are initiated from the vertical $y$-axis with $y = y_0$. In particular, this time we use the section $x = \dot{y} = 0$, $\dot{x} > 0$, launching orbits parallel to the $x$-axis. This allow us to construct again a 2D plane in which the $y$ coordinate of orbits is the abscissa, while the logarithmic value of the energy $\log_{10}(\widehat{C})$ is the ordinate. The orbital structure of the $(y,\widehat{C})$-plane when $\widehat{C} \in [0.001,1]$ is shown in Fig. \ref{xyc}b. The black solid line is the limiting curve which distinguishes between regions of allowed and forbidden motion and is defined as
\begin{equation}
f_L(y,\widehat{C}) = \Phi_{\rm t}(0,y) = E.
\label{zvc2}
\end{equation}
A very complicated orbital structure is reveled in the $(y,\widehat{C})$-plane which however, in general terms, is very similar with that of the $(x,\widehat{C})$-plane. One may observe that for $E > -1.37$ the forbidden regions of motion around $L_4$ and $L_5$ completely disappear. In fact the value $E(L_4) = E(L_5) = -1.37$ is another critical energy level. It is seen that above that critical value the left part $(y_0 < 0)$ of the plane is occupied almost completely with initial conditions of orbits that escape to sectors 1 or 4, while the right part of the same plane on the other hand, contains a mixture of initial conditions of escaping and non-escaping orbits.

It is evident from the results presented in Figs. \ref{xyc}(c-d) that the escape times of the orbits are strongly correlated to the escape basins. In addition, one may conclude that the smallest escape periods correspond to orbits with initial conditions inside the escape basins, while orbits initiated in the fractal regions of the planes or near the boundaries of stability islands have the highest escape rates. In both types of planes the escape times of orbits are significantly reduced with increasing energy. Combining all the numerical outcomes presented in Figs. \ref{xyc}(a-d) we may say that the key factor that determines and controls the escape times of the orbits is the value of the orbital energy (the higher the energy level the shorter the escape rates), while the fractality of the basin boundaries varies strongly both as a function of the energy and of the spatial variable. Indeed in 2D systems (see also \citet{EJSP08} and \citet{EP14}) the total energy is the key parameter which determines the escape times or orbits. In 3D systems on the other hand, the escape time of orbits is influenced not only by the total energy of the orbits but also by the spatial $z$ variable (see \citet{Z15}).

The following Fig. \ref{percs3}(a-b) shows the evolution of the percentages of the different types of orbits on the $(x,\widehat{C})$ and $(y,\widehat{C})$ planes as a function of the dimensionless energy parameter $\widehat{C}$. We see in Fig. \ref{percs3}a that for $\widehat{C} < 0.1$ the percentages of escaping orbits through channels 1 and 2 exhibit similar fluctuations, while for larger energy levels they start to diverge. In particular, the rate of escapers through exit 1 increases, while on the other hand the rate of escapers through exit 2 decreases. The portion of escaping orbits through sectors 3 and 4 is smaller, while in general terms the percentage of non-escaping regular orbits drops, although for some energy intervals the amount of regular orbits presents some peaks. In the sam vein one may observe in Fig. \ref{percs3}b that the percentages of escaping orbits through exits 1 and 2 identically evolve for $\widehat{C} < 0.01$. For higher values of the energy the rates diverge following completely different paths. We also see that each sector dominates for different range of the energy. It should be pointed out that the percentage of non-escaping ordered orbits exhibits a very smooth decrease for $\widehat{C} < 0.01$, while for larger energy levels several sudden peaks appear.

The evolution of the average value of the escape time $< t_{\rm esc} >$ of orbits as a function of the dimensionless energy parameter is given in Fig. \ref{stats}(a-b) for the $(x,\widehat{C})$ and $(y,\widehat{C})$ planes, respectively. It is seen, that for low values of energy, just above the escape value, the average escape time of orbits is about 1000 time units, however it reduces rapidly tending asymptotically to zero which refers to orbits that escape almost immediately from the system. We feel it is important to justify this behaviour of the escape time. As the value of the Jacobi constant increases the escape channels (which are of course symmetrical) become more and more wide and therefore, orbits need less and less time in order to find one of the openings in the open equipotential surface and eventually escape from the system. This geometrical feature explains why for low values of energy orbits consume large time periods wandering inside the open equipotential surface until they eventually locate one of the two exits and escape to infinity. Finally Fig. \ref{stats}(c-d) shows the evolution of the percentage of the total area $(A\%)$ on the $(x,\widehat{C})$ and $(y,\widehat{C})$ planes corresponding to basins of escape, as a function of the dimensionless energy parameter. It is seen that foe low values of the energy both types of planes are highly fractal. However, as we proceed to higher energy levels the degree of fractalization reduces and the area corresponding to basins of escape start to grow rapidly. Eventually, at very high energy levels $(\widehat{C} = 1)$ the fractal domains completely disappear and the well formed basins of escape occupy the entire planes.

\section{Discussion and conclusions}
\label{disc}

The aim of this work was to shed some light to the trapped or escaping character of orbits in a dynamical gravitational model describing a binary system of interacting dwarf spheroidal galaxies. As a first step we considered the case where the two galaxies are identical (same mass, same structure) similarly to the Copenhagen case of the classical restricted three-body problem. As far as we know, this is the first time that the escape process of stars in a binary system of interacting galaxies is systematically investigated in such detail.

We defined for several values of the Jacobi integral, dense uniform grids of $1024 \times 1024$ initial conditions $(x_0,y_0)$ and $(x_0, \dot{x_0})$ regularly distributed in the area allowed by the energy level on the configuration and phase space, respectively and then we identified regions of order/chaos and bound/escape. For each initial condition, the maximum time of the numerical integration was set to be equal to $10^4$ time units. However, when a particle escaped the numerical integration was effectively ended and proceeded to the next available initial condition. We managed to distinguish between ordered/chaotic and trapped/escaping orbits and we also located the basins of escape leading to different exit channels, finding correlations with the corresponding escape times of the orbits. Our numerical investigation suggests, that the overall escape mechanism is a very complicated procedure and very dependent on the value of the Jacobi integral.

The main numerical results of our research can be summarized as follows:
\begin{enumerate}
 \item In both the configuration and the phase space there are several stability islands which correspond to bounded motion around one or around both galaxies. In general terms we may say that around 10\% of orbits with initial conditions inside the Lagrangian radius are regular and therefore do not escape.
 \item At relatively high energy levels we found that in both the configuration and the phase space the majority of orbits escape through exit channel 1, while at low energy levels, just above the escape value, both exit channels are almost equiprobable.
 \item It was observed that in the 2D binary system of the interacting galaxies trapped chaotic orbits possess an extremely weak percentage (less than 0.01\% of the total integrated orbits). In 3D systems on the other hand (see \citet{Z15}) trapped chaos is prominent especially at values of energy above but very close to the escape energy.
 \item It was detected that the average escape time of orbits decreases almost exponentially with increasing energy mainly because simultaneously the width of the escape channels increases and therefore the orbits need less and less time to find one of the two openings and escape to infinity.
 \item Our calculations revealed that at low values of the energy the structure of all types of planes is highly fractal, while as we proceed to higher energy levels the degree of fractality decreases and several well-defined basins of escape emerge.
\end{enumerate}

Judging by the present outcomes we may say that our task has been successfully completed. We hope that the present numerical analysis and the corresponding results to be useful in the field of escape dynamics in binary galaxy systems. The outcomes as well as the conclusions of the present research are considered, as an initial effort and also as a promising step in the task of understanding the escape mechanism of orbits in systems of interacting galaxies. Taking into account that our results are encouraging, it is in our future plans to modify properly our dynamical model in order to expand our investigation into three dimensions and explore the entire six-dimensional phase space. Furthermore, we will consider more general cases where the two galaxies are not identical.

\section*{Acknowledgment}

The author would like to express his warmest thanks to the anonymous referee for the careful reading of the manuscript and for all the apt suggestions and comments which allowed us to improve both the quality and the clarity of the paper.


\begin{thebibliography}{}

\bibitem[\protect\citeauthoryear{Aguirre et al.}{2001}]{AVS01} Aguirre, J., Vallejo, J.C., Sanju\'{a}n, M.A.F 2001, Phys. Rev E, 64, 066208

\bibitem[\protect\citeauthoryear{Aguirre \& Sanju\'{a}n}{2003}]{AS03} Aguirre, J., Sanju\'{a}n, M.A.F. 2003, Phys. Rev. E, 67, 056201

\bibitem[\protect\citeauthoryear{Aguirre et al.}{2009}]{AVS09} Aguirre, J., Viana, R.L., Sanju\'{a}n, M.A.F. 2009, Rev. Mod. Phys., 81, 333

\bibitem[\protect\citeauthoryear{Barrio et al.}{2008}]{BBS08} Barrio, R., Blesa, F., Serrano, S. 2008, Europhys. Lett., 82, 10003

\bibitem[\protect\citeauthoryear{Benet et al.}{1996}]{BTS96} Benet, L., Trautmann, D., Seligman, T.H. 1996, CeMDA, 66, 203

\bibitem[\protect\citeauthoryear{Benet et al.}{1998}]{BST98} Benet L., Seligman, T., Trautman, D. 1998, CeMDA, 71, 167

\bibitem[\protect\citeauthoryear{Benettin et al.}{1980}]{BGGS80} Benettin, G., Galgani L., Giorgilli, A., Strelcyn, J.M. 1980, Meccanica 15, Part I: theory, 9; Part II: Numerical Applications, 21

\bibitem[\protect\citeauthoryear{Bleher et al.}{1988}]{BGOB88} Bleher, S., Grebogi, C., Ott, E., Brown, R. 1988, Phys. Rev. A, 38, 930

\bibitem[\protect\citeauthoryear{Bleher et al.}{1989}]{BOG89} Bleher S., Ott, E., Grebogi, C. 1989, Phys. Rev. Let., 63, 919

\bibitem[\protect\citeauthoryear{Bleher et al.}{1990}]{BGO90} Bleher, S., Grebogi, C., Ott, E. 1990, Physica D, 46, 87

\bibitem[\protect\citeauthoryear{Blesa et al.}{2012}]{BSBS12} Blesa, F., Seoane, J.M., Barrio, R., Sanju\'{a}n, M.A.F. 2012, Int. J. Bifurc. Chaos, 22, 1230010

\bibitem[\protect\citeauthoryear{Binney \& Tremaine}{2008}]{BT08} Binney, J., Tremaine, S. 2008, Galactic Dynamics, Princeton Univ. Press, Princeton, USA

\bibitem[\protect\citeauthoryear{Caranicolas \& Innanen}{2009}]{CI09} Caranicolas, N.D., Innanen, K.A. 2009, AN, 330, 20

\bibitem[\protect\citeauthoryear{Caranicolas \& Papadopoulos}{2009}]{CP09} Caranicolas, N.D., Papadopoulos, N.J. 2009, New Astronomy, 14, 207

\bibitem[\protect\citeauthoryear{Caranicolas \& Zotos}{2009}]{CZ09} Caranicolas, N.D., Zotos, E.E. 2009, MRC, 36, 875

\bibitem[\protect\citeauthoryear{Carpintero \& Aguilar}{1998}]{CA98} Carpintero, D.D., Aguilar, L.A. 1998, MNRAS, 298, 1

\bibitem[\protect\citeauthoryear{Carpintero et al.}{2014}]{CMD14} Carpintero, D.D., Maffione, N., Darriba, L. 2014, Astronomy and Computing, 5, 19

\bibitem[\protect\citeauthoryear{Churchill et al.}{1979}]{C79} Churchill, R.C., et al. 1979, in Como Conference Proceedings on Stochastic Behavior in Classical and Quantum Hamiltonian Systems, Volume 93, Lecture Notes in Physics, ed. G. Casati, J. Fords (Berlin: Springer)

\bibitem[\protect\citeauthoryear{Cincotta et al.}{2003}]{CGS03} Cincotta, P.M., Giordano, C.M., Sim\'{o}, C. 1003, Physica D, 182, 151

\bibitem[\protect\citeauthoryear{Contopoulos}{1990}]{C90} Contopoulos, G. 1990, A\&A, 231, 41

\bibitem[\protect\citeauthoryear{Contopoulos}{1990}]{C02} Contopoulos, G. 2002, Order and Chaos in Dynamical Astronomy. Springer, Berlin

\bibitem[\protect\citeauthoryear{Contopoulos \& Kaufmann}{1992}]{CK92} Contopoulos G., Kaufmann D. 1992, A\&A, 253, 379

\bibitem[\protect\citeauthoryear{Darriba et al.}{2012}]{DMCG12} Darriba, L.A., Maffione, N.P., Cincotta, P.M., Giordano, C.M. 2012, International Journal of Bifurcation and Chaos, 22, 1230033

\bibitem[\protect\citeauthoryear{de Assis \& Terra}{2014}]{dAT14} de Assis, S.C., Terra, M.O. 2014, CeMDA, 120, 105

\bibitem[\protect\citeauthoryear{de Moura \& Letelier}{2000}]{dML00} de Moura, A.P.S., Letelier, P.S. 2000, Phys. Rev. E, 62, 4784

\bibitem[\protect\citeauthoryear{Ernst et al.}{2008}]{EJSP08} Ernst, A., Just, A., Spurzem, R., Porth, O. 2008, MNRAS, 383, 897

\bibitem[\protect\citeauthoryear{Ernst \& Peters}{2014}]{EP14} Ernst, A., Peters, T. 2014, MNRAS, 443, 2579

\bibitem[\protect\citeauthoryear{Fouchard et al.}{2008}]{FLFF02} Fouchard, M., Lega, E., Froeschl\'{e}, Ch., Froeschl\'{e}, Cl. 2002, CeMDA, 83, 205

\bibitem[\protect\citeauthoryear{Froeschl\'{e} et al.}{1997}]{FGL97} Froeschl\'{e}, Cl., Gonczi, R., Lega, E. 1997, Planetary Space Science, 45, 881

\bibitem[\protect\citeauthoryear{H\'{e}non}{1969}]{H69} H\'{e}non, M. 1969, A\&A, 1, 223

\bibitem[\protect\citeauthoryear{Hut \& Bahcall}{1983}]{HB83} Hut, P., Bahcall, J.N. 1983, ApJ, 268, 319

\bibitem[\protect\citeauthoryear{Innanen}{1980}]{I80} Innanen, K.A. 1980, AJ, 85, 81

\bibitem[\protect\citeauthoryear{Jung}{1987}]{J87} Jung, C. 1987, J. Phys. A, 20, 1719

\bibitem[\protect\citeauthoryear{Jung et al.}{1999}]{JLS99} Jung, C., Lipp, C., Seligman, T.H. 1999, Ann. Phys., 275, 151

\bibitem[\protect\citeauthoryear{Jung et al.}{1995}]{JMS95} Jung, C., Mejia-Monasterio, C., Seligman, T.H. 1995, Phys. Lett. A, 198, 306

\bibitem[\protect\citeauthoryear{Jung \& Pott}{1989}]{JP89} Jung, C., Pott, S. 1989, J. Phys. A, 22, 2925

\bibitem[\protect\citeauthoryear{Jung \& Richter}{1990}]{JR90} Jung, C., Richter, P.H. 1990, J. Phys. A, 23, 2847

\bibitem[\protect\citeauthoryear{Jung \& Scholz}{1987}]{JS87} Jung, C., Scholz, H.J. 1987, J. Phys. A, 20, 3607

\bibitem[\protect\citeauthoryear{Jung \& Scholz}{1988}]{JS88} Jung, C., Scholz, H. 1988, J. Phys. A, 21, 3607

\bibitem[\protect\citeauthoryear{Jung \& Tel}{1991}]{JT91} Jung, C., Tel, T. 1991 J. Phys. A, 24, 2793

\bibitem[\protect\citeauthoryear{Kennedy \& Yorke}{1991}]{KY91} Kennedy, J., Yorke, J.A. 1991, Physica D, 51, 213

\bibitem[\protect\citeauthoryear{Lai et al.}{2000}]{LMG00} Lai, Y.-C., de Moura, A.P.S., Grebogi, C. 2000, Phys. Rev. E, 62, 6421

\bibitem[\protect\citeauthoryear{Lai et al.}{1993}]{LGB93} Lai, Y.-C., Grebogi, C., Bl\"{u}mel, R., Kan, I. 1993. Phys. Rev. Let., 71, 2212

\bibitem[\protect\citeauthoryear{Lau et al.}{1991}]{LFO91} Lau, Y.-T., Finn, J.M., Ott, E. 1991, Phys. Rev. Let., 66, 978

\bibitem[\protect\citeauthoryear{Lipp \& Jung}{1999}]{LJ99} Lipp, C., Jung, C. 1999, Chaos, 9, 706

\bibitem[\protect\citeauthoryear{Manos \& Athanassoula}{2011}]{MA11} Manos, T., Athanassoula, E. 2011, MNRAS, 415, 629

\bibitem[\protect\citeauthoryear{Martinez-Velpuesta \& Shlosman}{2004}]{MVS04} Martinez-Velpuesta, I., Shlosman, I. 2004, ApJ, 613, L29

\bibitem[\protect\citeauthoryear{Moser}{1958}]{M58} Moser, J. 1958, Commun. Pure Appl. Math., 11, 257

\bibitem[\protect\citeauthoryear{Motter \& Lai}{2002}]{ML02} Motter, A.E., Lai, Y.C. 2002, Phys. Rev. E, 65 R015205

\bibitem[\protect\citeauthoryear{Nagler}{2004}]{N04} Nagler, J. 2004, Physical Review E, 69, 066218

\bibitem[\protect\citeauthoryear{Nagler}{2005}]{N05} Nagler, J. 2005, Physical Review E, 71, 026227

\bibitem[\protect\citeauthoryear{Plummer}{1911}]{P11} Plummer, H.C. 1911, MNRAS, 71, 460

\bibitem[\protect\citeauthoryear{Poon et al.}{1996}]{PCOG96} Poon, L., Campos, J., Ott, E., Grebogi, C. 1996, Int. J. Bifurc. Chaos, 6, 251

\bibitem[\protect\citeauthoryear{Press et al.}{1992}]{PTVF92} Press, H.P., Teukolsky, S.A, Vetterling, W.T., Flannery, B.P. 1992, Numerical Recipes in FORTRAN 77, 2nd Ed., Cambridge Univ. Press, Cambridge, USA

\bibitem[\protect\citeauthoryear{S\'{a}ndor et al.}{2004}]{SESF04} S\'{a}ndor, Z., \'{E}rdi, B., Sz\'{e}ll, A., Funk, B. 2004, CeMDA, 90, 127

\bibitem[\protect\citeauthoryear{Seoane et al.}{2006}]{SASL06} Seoane, J.M., Aguirre, J., Sanju\'{a}n, M.A.F. Lai Y.C., 2006, Chaos, 16, 023101

\bibitem[\protect\citeauthoryear{Seoane et al.}{2007}]{SSL07} Seoane, J.M., Sanju\'{a}n, M.A.F., Lai, Y.C. 2007, Phys. Rev. E, 76, 016208

\bibitem[\protect\citeauthoryear{Seoane \& Sanju\'{a}n}{2008}]{SS08} Seoane, J.M., Sanju\'{a}n, M.A.F. 2008, Phys. Lett. A, 372, 110

\bibitem[\protect\citeauthoryear{Seoane et al.}{2009}]{SHSL09} Seoane, J.M., Huang, L., Sanju\'{a}n, M.A.F., Lai, Y.C. 2009, Phys. Rev. E, 79, 047202

\bibitem[\protect\citeauthoryear{Seoane \& Sanju\'{a}n}{2010}]{SS10} Seoane, J.M., Sanju\'{a}n, M.A.F. 2010, Int. J. Bifurc. Chaos, 9, 2783

\bibitem[\protect\citeauthoryear{Skokos}{2001}]{S01} Skokos, C. 2001, J. Phys. A: Math. Gen., 34, 10029

\bibitem[\protect\citeauthoryear{Skokos et al.}{2001}]{SBA07} Skokos, Ch., Bountis, T.C., Antonopoulos, Ch. 2007, Physica D. 231, 30

\bibitem[\protect\citeauthoryear{Zotos}{2012a}]{Z12a} Zotos, E.E. 2012a, PASA, 29, 161

\bibitem[\protect\citeauthoryear{Zotos}{2012b}]{Z12b} Zotos, E.E. 2012b, ApJ, 750, 56

\bibitem[\protect\citeauthoryear{Zotos}{2013}]{Z13} Zotos, E.E. 2013, PASA, 30, 12

\bibitem[\protect\citeauthoryear{Zotos}{2014a}]{Z14a} Zotos, E.E. 2014a, Nonlinear Dynamics, 76, 1301

\bibitem[\protect\citeauthoryear{Zotos}{2014b}]{Z14b} Zotos, E.E. 2014b, Nonlinear Dynamics, 78, 1389

\bibitem[\protect\citeauthoryear{Zotos}{2015}]{Z15} Zotos, E.E. 2015, MNRAS, 446, 770

\bibitem[\protect\citeauthoryear{Zotos \& Carpintero}{2001}]{ZC13} Zotos, E.E., Carpintero, D.D. 2013, CeMDA, 116, 417

\end{thebibliography}
\end{document}